**Emergence and dynamics of delusions and hallucinations across stages in early psychosis**

Catalina Mourgues-Codern, Ph.D.[1†], David Benrimoh, M.D.C.M.[2,3†], Jay Gandhi, B.A.[1], Emily A. Farina, Ph.D.[1], Raina Vin, B.S.[1], Tihare Zamorano, B.A.[2], Deven Parekh, M.Sc.[2], Ashok Malla, M.D.[2], Ridha Joober, M.D.[2], Martin Lepage, Ph.D.[2], Srividya N. Iyer, Ph.D.[2], Jean Addington, Ph.D.[4], Carrie E. Bearden, Ph.D.[5], Kristin S. Cadenhead, M.D.[6], Barbara Cornblatt, Ph.D.[7], Matcheri Keshavan, M.D.[8], William S. Stone, Ph.D.[8], Daniel H. Mathalon, M.D., Ph.D.[9], Diana O. Perkins, M.D., M.P.H.[10], Elaine F. Walker, Ph.D.[11], Tyrone D. Cannon, Ph.D.[1,12], Scott W. Woods, M.D.[1,13], Jai L. Shah, M.D.[2], Albert R. Powers, M.D., Ph.D.[1,10,12]*

1. Yale University School of Medicine and the Connecticut Mental Health Center, New Haven, CT, USA
2. McGill University, Montreal, Quebec, Canada
3. Stanford University School of Medicine, Palo Alto, CA, USA
4. University of Calgary, Calgary, Canada
5. University of California Los Angeles, Los Angeles, CA, USA
6. University of California San Diego, La Jolla, CA, USA
7. Hofstra Northwell School of Medicine, Uniondale, NY, USA
8. Beth Israel Deaconess Medical Center, Boston, Harvard Medical School, MA, USA
9. University of California San Francisco, San Francisco, CA, USA
10. University of North Carolina at Chapel Hill, Chapel Hill, NC, USA
11. Emory University, Atlanta, GA, USA
12. Yale University Department of Psychology, New Haven, CT, USA
13. Connecticut Mental Health Center, New Haven, CT, USA

*Correspondence should be addressed to:
Albert R. Powers, M.D., Ph.D.
The Connecticut Mental Health Center, Rm. S109
34 Park Street
New Haven, CT 06519
albert.powers@yale.edu
203.974.7329

†Contributed equally

**Running Title:** Delusions precede hallucinations across stages of early psychosis

**Words in main body of manuscript:** 2951

**Abstract**

Hallucinations and delusions are often grouped together within the positive symptoms of psychosis. However, recent evidence suggests they may be driven by distinct computational and neural mechanisms. Examining the time course of their emergence may provide insights into the relationship between these underlying mechanisms.

Participants from the second (N = 719) and third (N = 699) iterations of the North American Prodrome Longitudinal Study (NAPLS 2 and 3) were assessed for timing of CHR-P-level delusion and hallucination onset. Pre-onset symptom patterns in first-episode psychosis patients (FEP) from the Prevention and Early Intervention Program for Psychosis (PEPP-Montréal; N = 694) were also assessed. Symptom onset was determined at baseline assessment and the evolution of symptom patterns examined over 24 months.

In all three samples, participants were more likely to report the onset of delusion-spectrum symptoms prior to hallucination-spectrum symptoms (odds ratios (OR): NAPLS 2 = 4.09; NAPLS 3 = 4.14; PEPP, Z = 7.01, $P < 0.001$) and to present with only delusions compared to only hallucinations (OR: NAPLS 2 = 5.6; NAPLS 3 = 11.11; PEPP = 42.75). Re-emergence of delusions after remission was also more common than re-emergence of hallucinations ($Ps < 0.05$), and hallucinations more often resolved first ($Ps < 0.001$). In both CHR-P samples, ratings of delusional ideation fell with the onset of hallucinations ($P = 0.007$).

Delusions tend to emerge before hallucinations and may play a role in their development. Further work should examine the relationship between the mechanisms driving these symptoms and its utility for diagnosis and treatment.

**Introduction**

Hallucinations and delusions frequently co-occur in psychotic illness.[1–5] Since they were first described, these two symptoms have been grouped together into what are now known as the positive symptoms of psychosis.[6,7] However, emerging evidence suggests that delusions and hallucinations may be driven by distinct but related alterations in learning and inference.[8–13] Understanding the dynamic interrelationships between these symptoms could allow for a more nuanced understanding of the distinct mechanisms driving their emergence.

Historically, delusions have been characterized as either primary (e.g. emerging out of delusional mood, a state in which the environment seems to have changed in some unexplainable and potentially aversive way) or as secondary to some other phenomenon (e.g. a delusion arising out of a mood state or perceptual experience).[14] Broadly, cognitive models of psychosis describe delusion formation as the result of an individual's drive to make sense of a confusing or stressful experience.[15–18] One such model, the anomalous experience hypothesis, predicts that delusion formation is a response to unusual perceptual experiences or hallucinations and thus represents a (secondary) attempt by the individual to explain or understand their unusual perceptual experience.[19] In this view, delusions may be the product of normal reasoning processes used to explain confusing or potentially distressing experiences: a patient who hears a voice saying "Join the FBI" multiple times per day may be more likely to conclude and subsequently believe that the FBI is following her. While data exist that support this hypothesis and cognitive models of psychotic symptoms broadly,[3,20–27] these studies are somewhat limited by general-population sampling and cross-sectional designs.

Indirect evidence from the putatively prodromal Clinical High-Risk state for psychosis (CHR-P) indicates that hallucinations and delusions may not emerge at the same time: they co-occur much more frequently after the onset of threshold-level psychosis.[1,2,5] Additionally, delusional ideation is present in nearly all participants with perceptual abnormalities, but perceptual abnormalities are present in only some individuals with delusional ideation.[5,28] However, no known studies have directly examined the relative timing of symptom onset during and directly following symptom emergence. Ideally, longitudinal data would be obtained from large, naturalistically followed samples of individuals when positive symptoms are actively and acutely emerging. Doing so would allow for a more detailed and accurate understanding of symptom emergence and evolution within the natural history of psychosis.

We examined the onset and evolution of delusions and hallucinations over time in two large prospective CHR-P cohorts (N = 719 and 699) and one large retrospective (follow-back) first episode psychosis (FEP) cohort (N = 694). We investigated these questions across stages of illness in order to identify consistencies that may be present before and after psychosis onset.

**Methods**

*Participant groups*

All participants provided written informed consent after receiving a complete description of the constituent studies, which were approved by local ethics committees. This study included three main data sources: CHR-P participants recruited through the second and third waves of the multi-site North American Prodrome Longitudinal Study (NAPLS 2 and 3)[29,30] and an FEP sample from the Prevention and Early Intervention Program for Psychosis (PEPP-Montréal),[31] a Canadian specialized early intervention service. Further sample characteristics may be found in the Supplemental Materials.

NAPLS participants who met criteria for Attenuated Positive Symptom Syndrome (APSS) with CHR-P-level (scores 3-5) on delusional ideation (i.e., SOPS items P1 Unusual Thought Content or P2 Suspiciousness or Persecutory Ideas) and/or hallucinations (SOPS item P4 Perceptual abnormalities and Hallucinations) were included. The final CHR-P samples included 719 participants from NAPLS 2 and 699 from NAPLS 3.

The PEPP sample consisted of 694 consecutive patients in their first episode of psychosis who were deemed eligible and consented to participate. For eligibility, participants were required not to have taken antipsychotic medication for greater than 30 days prior to referral. Patients with IQ < 70, psychotic illness solely related to substance use, or non-psychiatric causes of mental disorder were not enrolled in PEPP.

*Measures*

*NAPLS 2 and 3*

The Structured Interview for Psychosis-Risk Syndromes (SIPS)[34] was administered to evaluate for a psychosis-risk syndrome by certified personnel at each site. For the purposes of brevity in this paper, we will discuss hallucination- and delusion-spectrum symptoms (e.g. attenuated psychotic symptoms) as hallucinations and delusions, respectively. As has been done previously,[20] CHR-P-level P1 (unusual thoughts) and P2 (suspiciousness) ratings (3-5) were taken as signifying the presence of early delusional ideation. P1 and P2 intensity has been shown to predict delusion formation[20] and progression of illness to psychosis.[28,35,36] CHR-P-level P4 (perceptual abnormalities) ratings (3-5) were taken as ratings of early hallucinatory phenomena. The maximum score for either P1 or P2 was used for subsequent analyses. Based on these ratings, participants were characterized as having only delusions, only hallucinations, both, or neither at each timepoint. Because the SIPS also elicits information on the timing of symptoms onset prior to baseline assessment, we were able to obtain information on the relative timing of onset of hallucinations and delusions even in those who presented with both symptoms at baseline. If the difference between the onset times of delusions and hallucinations was less than one month, we considered the participant to be in the combined-onset group. To further delineate participant trajectories, we identified those participants who ultimately entered remission from those who did not; participants with scores lower than 3 in P1, P2, and P4 during their last available visit were classified as remitters.

*PEPP*

Patients in the PEPP sample were assessed at baseline using the Circumstances of Onset and Relapse Schedule (CORS)[37], which allowed for retrospective determination of pre-onset (prodromal) and initial psychotic symptoms (the *first* prodromal symptom contiguous with psychosis onset), the *first* psychotic symptom, and the *first* psychotic symptom lasting longer than one week). For this analysis, and similar to CHR-P samples, CORS symptoms related to altered beliefs (odd beliefs, delusions, paranoia/suspiciousness) were considered on the delusion spectrum; altered perceptions and hallucinations were considered on the hallucination spectrum. In addition to the retrospective symptom reports, patients were prospectively assessed for psychotic symptoms using the Scale for the Assessment of Positive Symptoms (SAPS)[38] at intake and then at months 1, 2, 3, 6, 9, 12, 18 and 24, for a total of 12 timepoints. For the purposes of this study, the presence or absence of hallucinations and delusions in the prospective sample was established using SAPS questions 7 (global assessment of hallucinations) and 20 (global assessment of delusions); scores of 0 (none) and 1 (questionable) were coded as 'absent' and scores of 2 (mild) and above were coded as present. As not all subjects had data at each timepoint, analysis at each timepoint was based on the subjects who had provided data at that time.

Further details on measures collected are available in Supplemental Materials.

*Statistical Analysis*

Different pathways en route to a condition can influence the timing and variability of symptom expression[41]. We sought to determine if hallucinations and delusions exhibited different patterns of variability over time, which could support the possibility of different pathways toward symptom expression. First, we analyzed variability in the severity scores of delusions and hallucinations across subjects over time. The standard deviation of individual symptom scores over time was computed and then compared using a t-test for independent samples (**Fig. 2a,d,g**). Next, we examined variability in the pattern of occurrence of delusions and hallucinations separately over time (regardless of severity) using Lempel-Ziv complexity (LZC)[42]; see Supplemental Materials for details. Because our objective was to compare variability of symptom presence over time, we excluded participants with fewer than 3 timepoints and those who never experienced the symptoms of interest. Mean LZC, computed separately for delusions and hallucinations, was compared using independent-samples t-tests (**Fig. 2b,e,h; Table S23**). Finally, after analyzing variability both in the graded severity and binary pattern of occurrence of each symptom separately, we examined the probability of transitioning between symptom states (having delusions, hallucinations, neither, or both) over time (**Fig. 2c,f,i**). Symptom states were defined by the presence or absence of each individual symptom.

If onset of hallucination emergence is a marker of worsening clinical condition in those who already have delusions, we would expect that delusion severity would also increase along with hallucination onset. Alternatively, if the mechanisms driving hallucinogenesis are a compensatory response to the mechanisms driving delusion formation, we might expect a decrease in delusion severity with hallucination onset. We examined differences in symptom severity across symptom onset timepoints by aligning to hallucination onset and testing for differences across that timepoint using pairwise t-tests.

**Results**

***Delusions emerge before hallucinations in most CHR-P and FEP participants.***

In all samples studied, participants were more likely to present at baseline assessment with only delusions than with only hallucinations (**Fig. 1, Tables S2, S6 and S11;** NAPLS 2: OR = 5.6, CI$_{95}$ 3.52 - 7.27; NAPLS 3: OR = 11.11, CI$_{95}$: 6.97 - 17.71; PEPP: OR = 42.75, CI$_{95}$: 18.81 - 97.19)**.** On examination of retrospective reports, delusions also occurred as a first symptom more frequently than hallucinations across all samples (**Fig. 1,** dotted line; NAPLS 2: OR = 4.09, CI$_{95}$ 3.20 - 5.23; NAPLS 3 OR = 4.14, CI$_{95}$ 3.24-5.28; PEPP Z = 7.01, P < 0.001 (first prodromal symptom) = 4.04, CI$_{95}$ 2.77 - 5.90). Elapsed time between first and second symptom onset did not differ significantly between groups (see **Supplemental Results, Table S22**).

***Hallucinations are more volatile than delusions in CHR-P.***

As shown in **Figure 1**, participants' symptom profiles evolved in both prospective datasets, with symptoms appearing, resolving, and re-emerging over time (**Tables S18 to S20**). We investigated these patterns in all samples (**Fig. 2**).

We first examined the variance in symptom scores over time for delusions and hallucinations. Hallucinations exhibited a higher mean standard deviation over time than delusions in both CHR-P samples (**Fig. 2a,d**; NAPLS 2: $T_{718}$ = 4.836, NAPLS 3: $T_{559}$ = 3.397, $Ps$ < 0.001). Delusions exhibited a higher mean standard deviation over time in the first-episode psychosis sample (**Fig. 2g**; PEPP: $T_{1069}$ = 2.884, $P$ = 0.004); this may have been influenced by a floor effect in hallucination ratings in this sample (**Fig. S3**).

Next, we examined the complexity of patterns of delusion and hallucination occurrence–how much each symptom appeared and disappeared over time, measured by Lempel-Ziv Complexity (LZC). We found that the LZC of the symptom expression pattern was significantly higher in hallucinations compared to delusions in both CHR-P samples (**Fig. 2b**, NAPLS 2: $T_{718}$ = 2.175, $P$ = 0.029; **Fig. 2e**, NAPLS 3 $T_{559}$ = 2.113, $P$ = 0.035) but not in the first-episode sample ($T_{1069}$ = 0.6118, $P$ = 0.0541, **Fig. 2h**, **Table S23**).

*Hallucinations tend to re-emerge after re-emergence of delusions and resolve more quickly.*

Initial emergence of hallucinations and delusions can only be observed once per participant. However, because participants in all samples were followed over time, it was possible to determine whether patterns of symptom re-emergence across the whole sample recapitulated initial patterns of emergence.

The order of re-emergence and resolution of symptoms over time echoed patterns seen on initial symptom emergence. In all samples' prospective data, participants were more likely to transition from a state of having no symptoms to having delusions alone compared to hallucinations alone (**Fig. 2c**; $Z_{NAPLS\,2}$ = 2.2692, $P$ = 0.023, **Fig. 2f**; $Z_{NAPLS\,3}$ = 3.008, $P$ = 0.003, **Fig. 2i**; $Z_{PEPP}$ = 9.9216, $P$ < 0.001, **Table 21**). In the opposite direction, hallucination resolution was also more likely than delusion resolution from a state of having both delusions and hallucinations ($Z_{NAPLS\,2}$ =6.7021, $Z_{NAPS\,3}$ = 7.0841, $Z_{PEPP}$ = 5.0529, $Ps$ < 0.001).

*Delusion severity falls after hallucination onset.*

We next examined the relationship between symptoms at time of initial hallucination onset among CHR-P individuals who developed hallucinations while being followed prospectively (**Fig. 3**).

The initial emergence of hallucinations was found to be preceded by a period of supra-threshold delusion intensity, followed by a decrease in both delusion and hallucination intensity after hallucination onset in both the NAPLS 2 (**Fig. 3a**) and NAPLS 3 (**Fig. 3b**) samples (combined NAPLS 2 and NAPLS 3 samples, pairwise $T_{43}$ = 2.831, $P$ = 0.007). This pattern was observed in those whose symptoms ultimately remitted ($T_{15}$ = 3.727, $P$ = 0.002) but not in those whose symptoms persisted or worsened over time ($T_{24}$ = .492, $P$ = 0.627; **Fig. S1**).

**Discussion**

Examining the temporal order of psychotic symptom emergence has the potential to deepen our understanding of both phenomenology and the mechanisms driving those symptoms. Consistent patterns of emergence and resolution may help to establish a pathophysiological model for symptom development within psychosis,[43,44] ultimately shifting the focus of intervention from symptom alleviation to interruption of the pathophysiological processes leading to disease.[45–48]

Across three large samples of individuals in the earliest phases of psychosis, each using retrospective and prospective approaches to symptom measurement, delusions emerge before hallucinations in the majority of those studied. Delusions similarly re-emerge before hallucinations after initial symptoms have remitted, recapitulating the patterns of initial emergence. Hallucinations exhibit a noisier symptom profile, with later emergence and higher symptom volatility over the CHR-P period. Lastly, rather than increasing following hallucination onset, the severity of delusions decreases with hallucination onset in both prospective CHR-P samples. Together, these results provide converging evidence that hallucination emergence may be a secondary (i.e., compensatory) response to the factors driving delusion formation in a majority of individuals in the earliest phases of psychosis.

While perhaps surprising, this possibility is consistent with emerging work in computational psychiatry, which has related the constituent positive symptoms to distinct and opposing alterations in learning and perceptual inference:[10,49,50] delusions may initially form as the result of inappropriate learning spurred by aberrant bottom-up noise driving experiences of aberrant salience or prediction-error signaling,[27,50–54] whereas hallucinations have consistently been associated with a top-down dominance of expectations (or priors) in perception,[49,55–57] which also tracks with symptom severity over time.[58] Although these mechanisms appear to be in contradiction, we[9] and others[10,59] have proposed that they may simply be separated in time: hallucinations and top-down dominance may be the result of processes attempting to compensate for the pre-existing bottom-up noise that first drives delusion formation. In this rendering, delusions would be expected to emerge before hallucinations in the majority of those studied, as we find here. The unexpected finding that delusion severity falls with hallucination onset is also consistent with this account, potentially reflecting an adaptive downweighting of noisy sensory information or a decrease in delusion-related distress with increased confidence in existing beliefs[60].

The present findings may also hint at dynamics accompanying transition to a first psychotic episode. Variance in hallucination expression over the CHR-P period is higher than that of delusions. This generally supports the hypothesis that hallucinations are the result of secondary processes: assuming both symptoms are the result of noisy neurobiological processes, the noise inherent in the pathogenesis of two (rather than one) pathophysiological processes may be expected to be additive.[61] It is notable that this difference is not present in the FEP sample, in which we see comparable variability of expression patterns in the two symptoms (**Fig. 2h**). Within our framework, this may denote the development of a steady state between the processes driving delusions and hallucinations, but further work is required to test this idea. Despite these differences, the probabilities associated with symptom emergence, remission, and re-emergence remain the same: delusions emerge and re-emerge before hallucinations. The consistency of this pattern may reflect consistency in the pathophysiological processes driving symptom formation, which is ultimately the result of the ways these underlying processes dynamically interact.

Despite the overall patterns, there is appreciable heterogeneity in symptom onset across all samples. Some individuals develop hallucinations prior to delusions and present for baseline ascertainment with only hallucinations present. As noted in **Tables S1 and S12-14**, these individuals may also exhibit distinct clinical features. Most notably, they are younger and have milder illness and higher functioning. Identification of processes that drive differential patterns of symptom emergence and how they might relate to general factors that influence functioning should be a focus of future work. The observational nature of the PEPP sample limits our ability to evaluate the impact of antipsychotic medications in that data set, and antipsychotics were not a factor in the NAPLS sample by design. The role of medications should also be better delineated in future work. Therapeutic implications of these findings may extend beyond antipsychotic use: if incoming noise that drives delusion formation is a primary abnormality in most people with psychosis, medications and psychotherapies meant to correct or cope with that abnormality before hallucinogenesis may be particularly promising[62].

One strength of the current work lies in its use of large samples across the phases of early psychosis. Because there are substantive differences between CHR-P and FEP sample characteristics,[63] convergent evidence is particularly valuable. There are also significant limitations to our efforts. There is inherent difficulty in using broad, clinically-oriented instruments to relate to specific, fine-grained neural processes driving symptom development, despite the repeatedly-demonstrated utility of these instruments for predicting clinical course.[36,64,65] For example, the SOPS P4 item includes low-level perceptual distortions in addition to hallucinations; although this inclusion makes it even more remarkable that we see delusions emerging before hallucinations across samples, more specificity would be desirable. New consensus instruments[66] that separately rate distress, interference, and tenacity ratings could aid future analyses. Ongoing efforts to gather detailed phenotypic and neural data over time as psychosis develops offer an unprecedented opportunity to replicate our findings and test the hypotheses that arise from them.[67]

**Figure Legends**

**Figure 1. Delusions emerge before hallucinations in most CHR-P and FEP participants.** Sankey diagrams of prevalence of attenuated psychotic symptoms at emergence, baseline assessment, and follow-up timepoints in NAPLS 2 (**a**, N = 719), NAPLS 3 (**b**, N = 699), and PEPP (**c**, N = 694). Participants were 4.15, 8.52, and 3.36 times more likely to report delusions as a first symptom than hallucinations, respectively. The width of the edges between nodes is proportional to the percentage of participants that transition from one symptom to another. Dotted line on the x axis denotes retrospectively-obtained data. Red = hallucinations; blue = delusions; gray = remission. See Tables S18 - S20 for raw category and transition numbers.

**Figure 2. Delusions are more stable and re-emerge before hallucinations.** Over the time followed, hallucination severity scores (**a,d**) were more variable in the CHR-P samples, and hallucinations were much more likely to be volatile in their expression, appearing and resolving more frequently than delusions, which were more stable (**b,e**). These findings were not present in the FEP sample (**g,h**). However, in all three samples, the probability of transitioning from no CHR-P level symptoms to CHR-P level delusions alone was significantly higher than the probability of transitioning to hallucinations alone (**c,f,I**). Hallucination resolution was also more likely than delusion resolution from a state of having both CHR-P level delusions and hallucinations. Red = hallucinations; blue = delusions; gray = remission; ** $p < 0.01$; *** $p < 0.001$. Line shading and error bars = 1 SEM.

**Figure 3. Delusion severity falls after hallucination onset.** In both NAPLS 2 (**a**) and NAPLS 3 (**b**), delusion severity on the SOPS decreased after hallucination onset (time 0). Combined NAPLS 2 and NAPLS 3 samples, pairwise $T_{43} = 2.831$, $P = 0.007$. Red = hallucinations; blue = delusions.

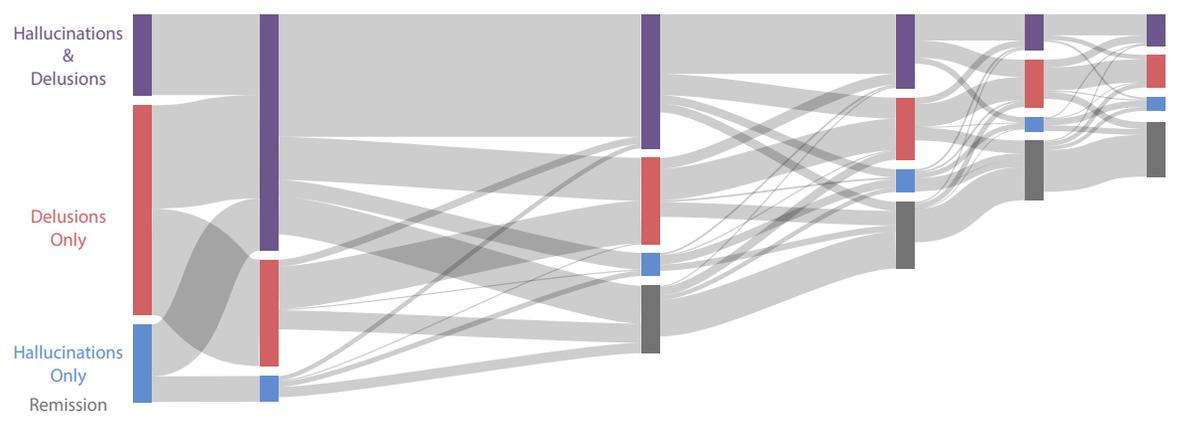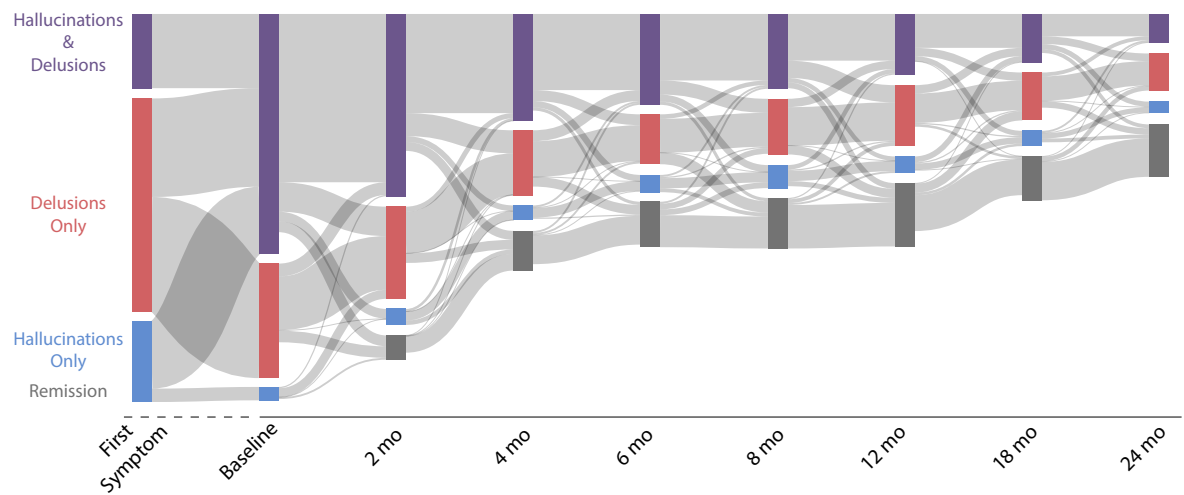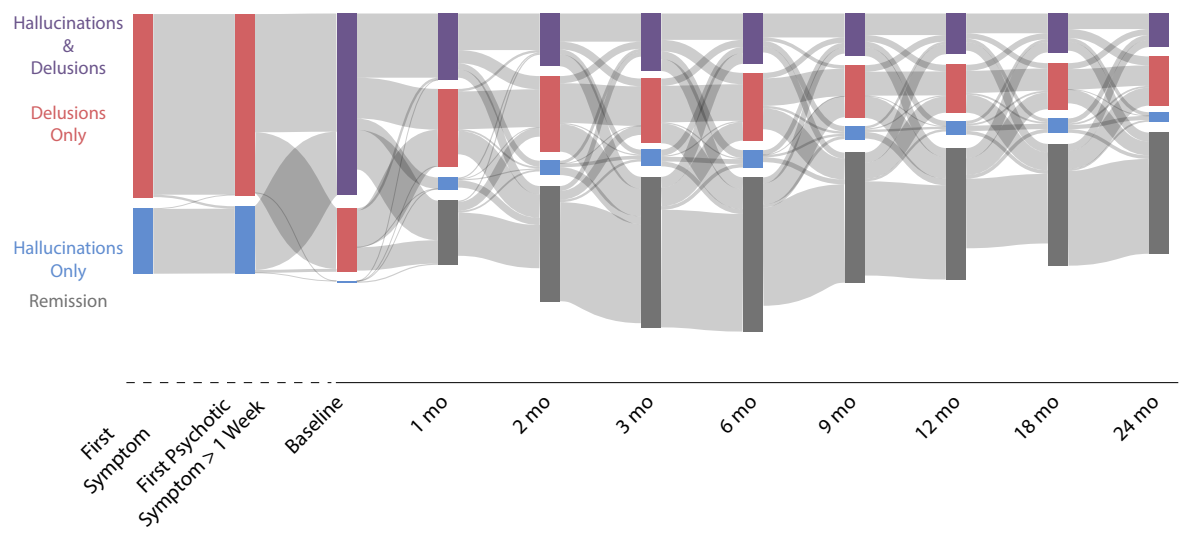

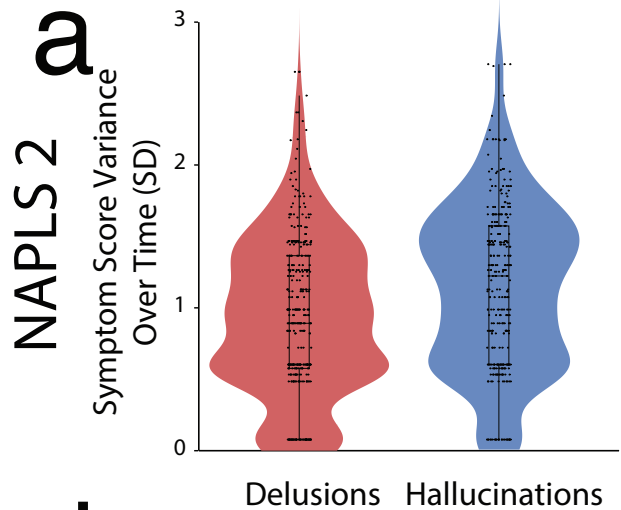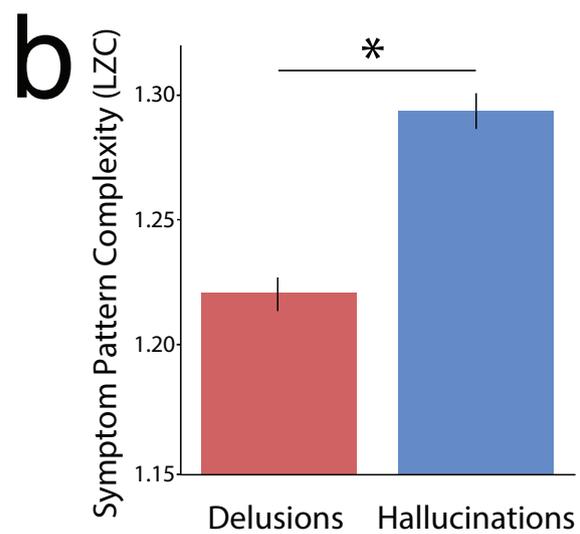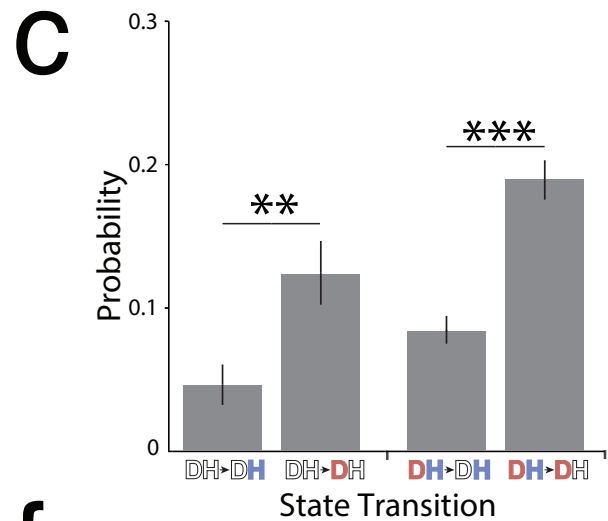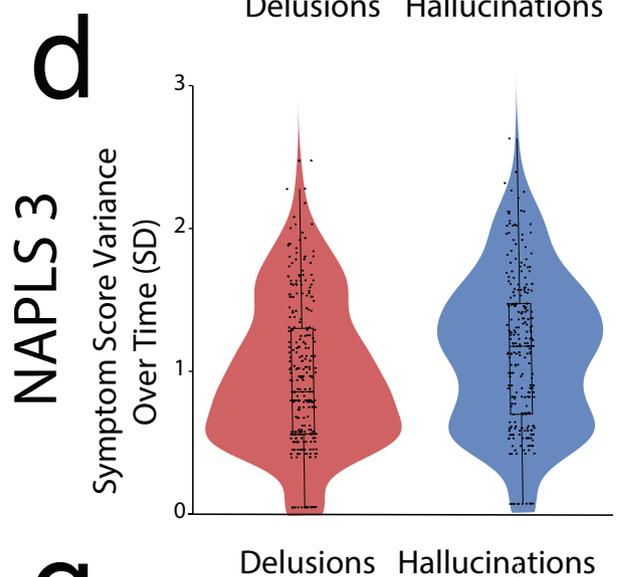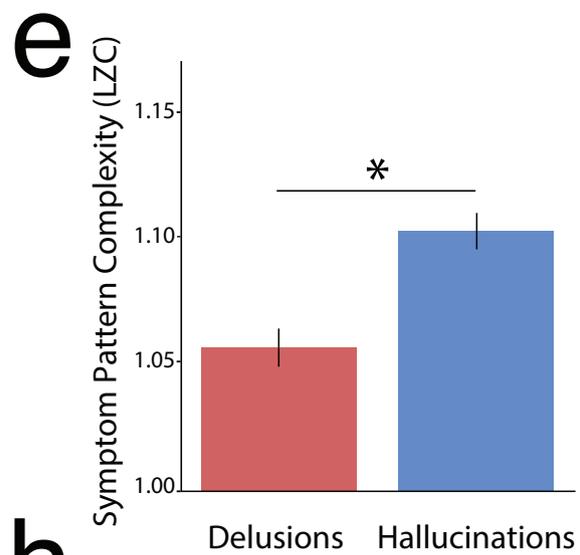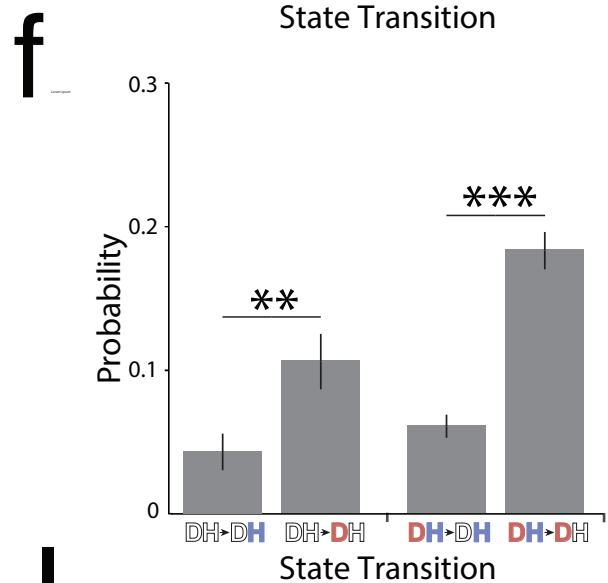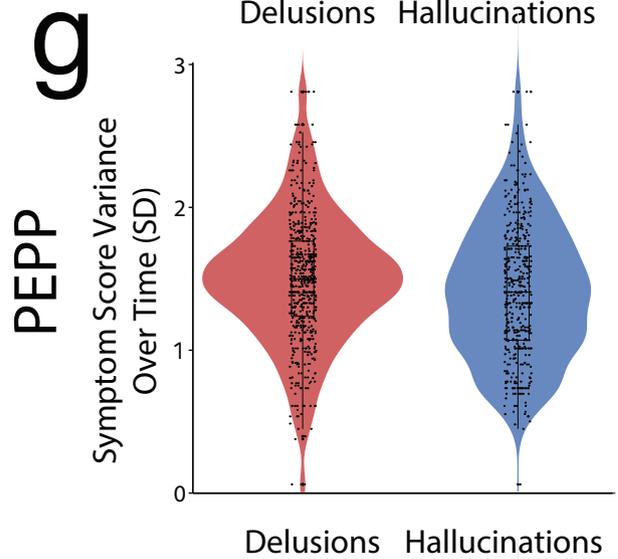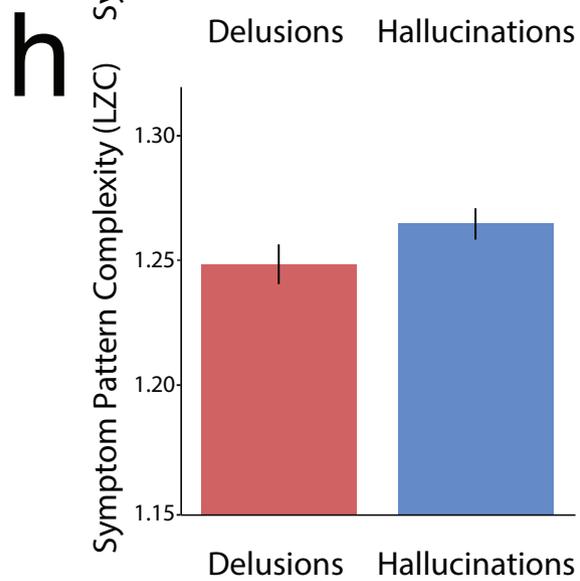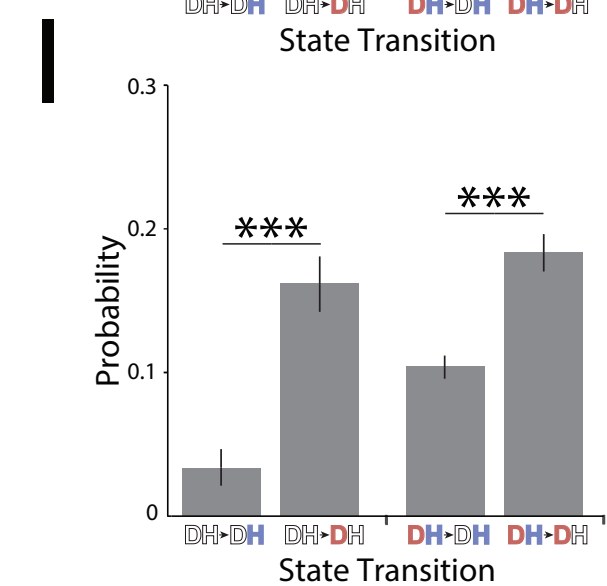

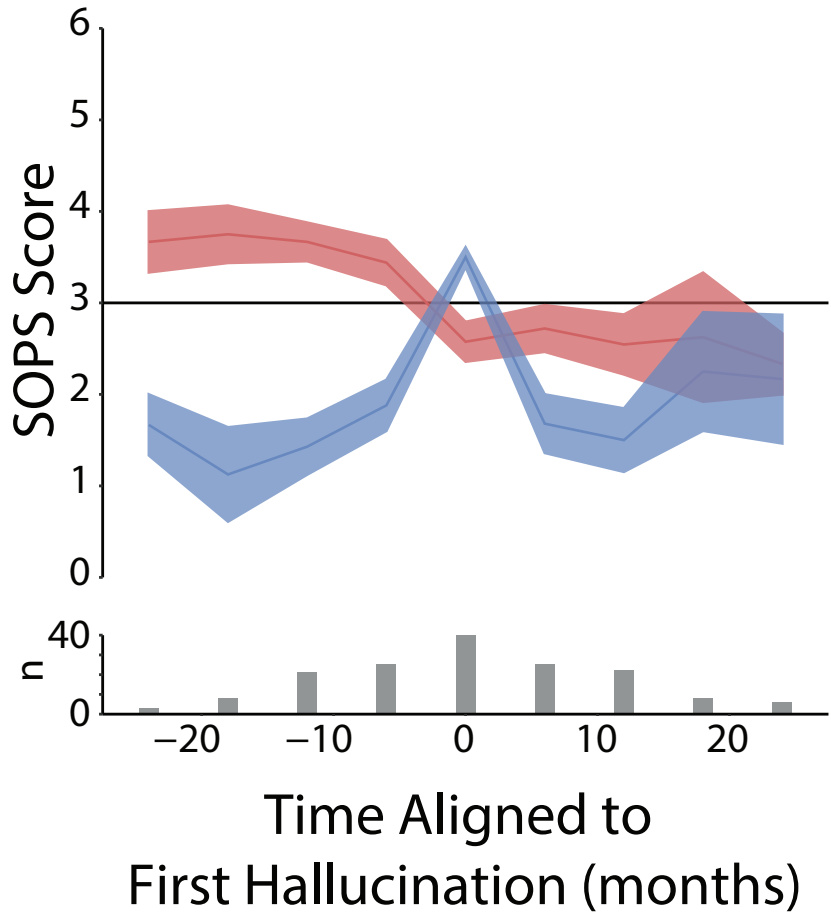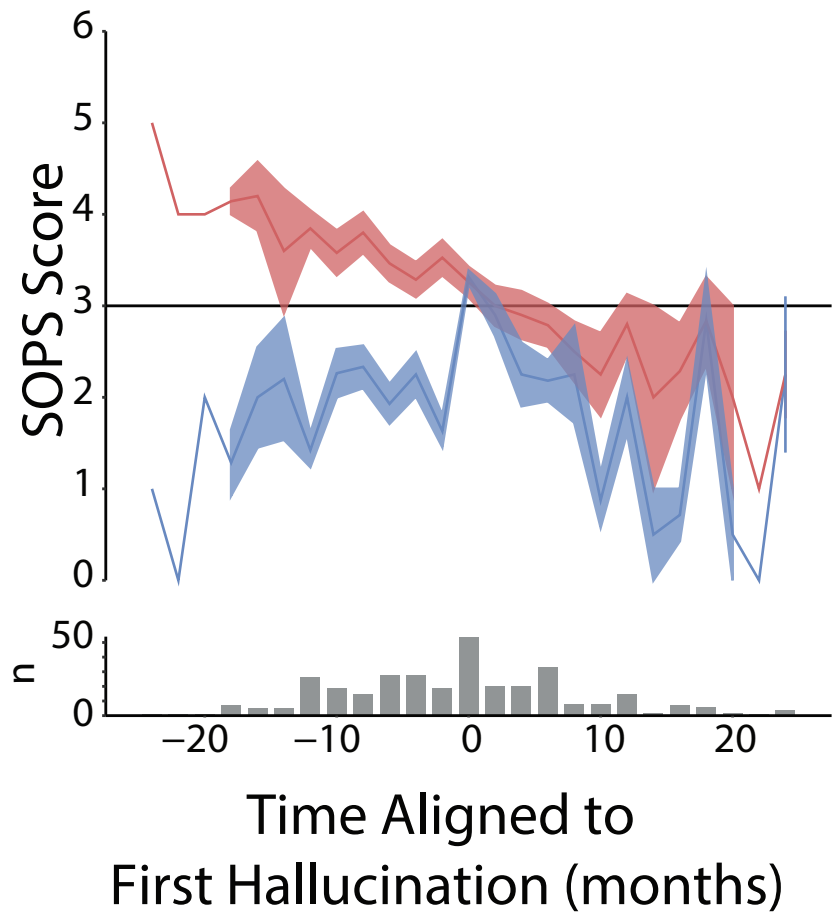

# Supplementary Information

# Emergence and dynamics of delusions and hallucinations across stages in early psychosis

**Supplemental Methods**

*Sample*

CHR-P participants met the Criteria of Psychosis-Risk Syndromes (COPS) based on the Structured Interview for Psychosis-Risk Syndromes (SIPS).[1] FEP participants from PEPP-Montréal[2,3] (hereinafter referred to as PEPP) were admitted to the service for a first episode of affective or non-affective psychosis between 2003 and 2018.

NAPLS 2 was conducted from 2008 to 2013. Participants were between 12 and 35 years old, and 764 met inclusion criteria for CHR-P. NAPLS 3 was conducted from 2014 to 2019. Participants were between 12 and 30 years old, and 710 met diagnostic criteria for a prodromal syndrome rated on the Scale of Prodromal Symptoms (SOPS). See Addington et al. for more details about the study and sample. [4,5] From NAPLS 2, 45 participants were excluded: 25 were below CHR-P level for these symptoms and 20 scored in the psychotic-level range. From the NAPLS 3 sample, 11 were excluded: 6 scored below CHR-level for these symptoms, and 5 scored in the psychotic-level range and were designated as having Brief Intermittent Psychotic Syndrome (BIPS).

The PEPP program is publicly funded and serves a catchment area of ~300,000 people; patients are drawn from multiple referral sources, including self-referral. In the PEPP dataset, patients were between the ages of 14 and 35 at time of referral. The selected sample consisted of 694 patients at baseline. PEPP serves as the sole first episode psychosis service in this area. The overall PEPP study received approval from the Douglas Hospital Research Centre's research ethics board, and participants provided written informed consent to be enrolled in the PEPP cohort. 80.51% of the patients accepted into the PEPP protocol study during this time period accepted to participate in the study.

*Measures*

*NAPLS 2 and 3*

In both NAPLS 2 and 3, participants were administered various tasks, interviews, and self-report measures. Participants from NAPLS 2 were assessed at baseline and four follow-ups over two years: 6, 12, 18, and 24 months. Participants from NAPLS 3 were also assessed over two years, with more frequent follow-ups: every two months in the first year and then at 18 and 24 months.

The Structured Interview for Psychosis-Risk Syndromes (SIPS)[6] was administered to evaluate for a psychosis-risk syndrome Presence of CHR-P corresponded to: 1) attenuated positive symptoms (APS) or fully psychotic positive symptoms occurring within a very brief period of time; and/or 2) decline in global functioning with schizotypal personality disorder earlier than age 19; and/or 3) a family history of schizophrenia-spectrum illness accompanied by a decline in functioning. The final sample consisted of only individuals meeting criteria for APS. The SIPS is accompanied by an instrument, the Scale of Prodromal Symptoms (SOPS), which assists in the rating of the severity of relevant symptoms on a 7-

point scale ranging from absent (scores from 0 to 2), moderate to severe but not psychotic (CHR-P level, 3 to 5) and severe and psychotic (score of 6). The SOPS gauges several distinct categories of prodromal symptom domains including positive (unusual thoughts, suspiciousness, grandiosity, perceptual abnormalities, disorganized communication) and negative dimensions (social anhedonia, avolition, expression of emotion, experience of emotions and self, ideational richness, occupational functioning). Ratings of positive symptoms in the range of 3 to 5 are required for designation as being at CHR-P.

As has been done previously,[7] P1 (unusual thoughts) and P2 (suspiciousness) ratings were taken as signifying the presence of early delusional ideation. P1 and P2 intensity has been shown to predict delusion formation[7] and progression of illness to psychosis.[8–10] P4 (perceptual abnormalities) ratings were taken as ratings of early hallucinatory phenomena. The maximum score for either P1 or P2 was used for subsequent analyses. Those with P1 or P2 scores between 3 and 5 at baseline who did not meet criteria for P4 were classified as participants with only delusions. Participants with P4 scores between 3 and 5 at baseline who did not meet criteria on P1 or P2 were categorized as those with only hallucinations. During the analysis, we did not consider grandiose ideas (P3). We made this decision based on the following factors: a) Grandiose ideas occur less frequently than P1 and P2 at baseline (symptoms at baseline in the range of 3 to 5, NAPLS 2, P1 = 82.3%, P2= 67.5%, P3 = 16.3%, NAPLS 3, P1 = 89.99%, P2 = 72.96% and P3 = 17.88%) and they have a significant but not complete overlap with P1, P2, and P4 (96.58% in NAPLS 2 and 71, 35% NAPLS 3, P3 co-occur with P1 or P2, 13.2% in NAPLS 2 and 19.2% in NAPLS 3 co-occur with P4). b) According to factorial analysis, P3 usually loads on a different factor than P1 and P2[11,12]. This indicates that there may be a different delusional process involved in P3 than in P1 and P2. c) Lastly, grandiose ideas have not been found to predict psychosis strongly and are not coded separately in the PEPP sample using the CORS. Participants who had SOPS scores between 3 and 5 on both P4 and either P1 or P2 were categorized as having both delusions and hallucinations. Because the SIPS also elicits information on the timing of onset of symptoms that occurred at (and prior to) the baseline assessment, we were able to obtain information on the relative timing of onset of hallucinations and delusions even in those who presented with both symptoms at baseline. Thus, we were able to identify two additional groups, those who started with hallucinations and then developed delusions and those who developed delusions and then hallucinations. If the difference between the onset times of delusions and hallucinations was less than one month, we considered the participant to be in the combined-onset group. To further delineate participant trajectories, we identified those participants who ultimately entered remission from those who did not; participants with scores lower than 3 in P1, P2, and P4 during their last available visit were classified as remitters.

*PEPP*

Patients in the PEPP sample were assessed at baseline using the Circumstances of Onset and Relapse Schedule (CORS)[13] which allowed for a retrospective determination of prodromal and initial psychotic symptoms (the *first* prodromal symptom, the *first* psychotic symptom, and the *first* psychotic symptom lasting longer than one week). For this analysis, and similarly to what was done in the CHR-P samples, CORS symptoms related to altered beliefs (odd beliefs, delusions, paranoia) were considered on the delusion spectrum; altered perceptions and hallucinations were considered on the hallucination spectrum. Note that the data available for the retrospective timepoints differs from the data collected at later timepoints: the retrospective data asks about the *first* symptom in a given category, and definitive information about *concurrent* symptoms is not available.

In addition to the 3 retrospective timepoints, patients were prospectively assessed for psychotic symptoms using the Scale for the Assessment of Positive Symptoms (SAPS)[14] at intake and then at months 1, 2, 3, 6, 9, 12, 18, and 24, for a total of 12 timepoints. For the purposes of this study, the presence or absence of hallucinations and delusions in the prospective sample was established using the SAPS questions 7 (global assessment of hallucinations) and 20 (global assessment of delusions); scores of 0 (none) and 1 (questionable) were coded as 'absent' and scores of 2 (mild) and above were coded as present. While remission of symptoms on the SAPS is often considered to be present at mild symptoms and below,[15] our choice for binarization was made as we were interested in following subtle changes in symptoms over time that might reflect underlying computational processes. In addition, as this was a treated sample, examining mild symptoms allowed for the assessment of symptoms that were present though symptomatically controlled by medication, which is likely relevant to understanding symptom transitions over time. As not all subjects had data at each point, analysis at each point was based on the subjects who had provided data at that time.

*Statistical Analysis*

*R studio* (RStudio, PBC, Boston, MA) and Python were used for all statistical analyses and visualizations, utilizing packages *networkD3, longCatEDA,* and *Matplotlib*. We constructed a Sankey diagram to depict the flow of patients to and from our established symptom categories over time using group transition frequencies at each timepoint (**Fig. 1**). To aid comparison with the other groups, the Sankey diagram for the PEPP sample (**Fig. 1c**) only included participants with a first reported prodromal symptom of delusions or hallucinations. However, the PEPP sample also included participants with other types of first symptoms (e.g., disorganization or mood related), and these participants were retained for subsequent analyses[16,17]. Proportions of participants reporting symptoms of each category as first symptoms were compared using odds ratios, chi-squared tests, z-tests, and post-hoc tests adjusted for multiple comparisons when applicable. Differences in clinical features were analyzed using ANOVA.

In the NAPLS 2 and 3 samples, the time elapsed between the first and second symptoms was calculated by subtracting the onset dates of the first and second symptoms. The mean onset times per symptom were compared using an independent samples t-test (**Table S22**).

We calculated LZC based on the binary time course of delusion and hallucination presence, considering SIPS or SAPS score > 2, as above. LZC is a measure of variability of a binary sequence; small numbers connote fewer observed patterns, while higher numbers connote greater variability in the observed patterns. The probability of discovering additional patterns in a sample is correlated with the number of measurement points taken. As our samples consist of varying measurement time points, we calculated standardized LZC scores to facilitate easy comparison between samples. Missing data points were not considered as patterns in our analysis.

**Supplemental Results**

*Symptom presentation at Baseline*

In NAPLS 2, 95.3% of the participants met the criteria for Attenuated Psychotic Symptoms, while in NAPLS 3, the percentage was 99.4%. The remaining percentages corresponded to Genetic Risk and Functional Decline psychosis-risk syndrome.

In NAPLS 2, for those that presented with delusions only, the onset of hallucinations was, on average, 11.35 months (SD = 5.66, n =28) after baseline. The onset of delusions for those who presented only hallucinations at baseline was reported on average at 10.33 months (SD = 8.24, n = 12) from baseline. In the NAPLS 3 sample, those who reported only delusions, the onset of hallucinations was reported on average 6.02 months (SD = 5.61, n = 47) after baseline, and for those with only hallucinations, the onset of delusions was reported at an average of 8.71 months (SD = 6.10, n = 7) after baseline. In the PEPP sample, the mean prodrome length was 22.92 months (SD= 38.09, n = 541; **Table S22**).

Across all three samples, mean SOPS scores for hallucinations were lower than those for delusions (**Fig. S3**; NAPLS 2, $M_{Hall}$ = 2.56, $SD_{Hall}$ = 1.024, $SEM_{Hall}$ = 0.063, $M_{Del}$ = 3.10, $SD_{Del}$ = 0.977, $SEM_{Del}$ = 0.05, $T_{718}$ = 6.4597, $P < 0.001$; NAPLS 3, $M_{Hall}$ = 2.524, $SD_{Hall}$ = 1.064, $SEM_{Hall}$ = 0.057, $M_{Del}$ = 3.01, $SD_{Del}$ = 0.967, $SEM_{Del}$ = 0.049, $T_{559}$ = 6.3997, $P < 0.001$; PEPP, $M_{Hall}$ = 1.15, $SD_{Hall}$ = 1.08, $SEM_{Hall}$ = 0.04, $M_{Del}$ = 1.75, $SD_{Del}$ =0.912, $SEM_{Del}$ = 0.037, $T_{1203}$ = 10.2873, $P < 0.001$)

*Symptom progression*

In NAPLS 2, approximately ~50% of participants reported the same symptoms between two consecutive visits. However, the percentage of participants who reported hallucinations over two consecutive visits was lower than the other symptoms. Once the symptoms disappear, there is a higher likelihood of remaining without symptoms than having a new one on the next visit. Only a small percentage of the symptoms reemerge, with delusions only being the highest proportion of symptoms reappearance, followed by delusions and hallucinations and hallucinations only. Hallucinations were the most frequent symptom before the symptom disappeared. The patterns were similar in NAPLS 3. The highest percentages corresponded to participants who, from one visit to the next one, had the same symptoms. Most participants who reported no symptoms in a visit remained with no symptoms to the next one. Unlike NAPLS 2, in this sample, when the symptoms reemerge, delusions only and hallucinations only proportionally more frequent than hallucinations and delusions. The same was true when participants transitioned from not having symptoms to having new ones (see **Tables S18** to **S20** for transition frequencies and conditional probabilities).

In the PEPP sample, patients with neither symptom continue to have no symptoms at the next time point; when symptoms recur, delusions are the most frequent category, followed by delusions and hallucinations, and finally by hallucinations alone. Those with delusions only commonly persist in having delusions at each transition point, with transitions to delusions and hallucinations becoming symptom-free being more likely than transition to hallucinations only. Similarly, those with hallucinations only do not tend to transition to only delusions- but rather to continue having hallucinations (at lower rates than those with delusions continue having delusions only), to switch to hallucinations and delusions, or to become symptom free.

*Clinical characteristics of symptom groupings*

Clinical characteristics also varied among symptom emergence profiles (**Table S12 to S15**). Participants who presented with only hallucinations at baseline were younger than the other groups (NAPLS 2, $F_{2, 716} = 11.24$, p < 0.001; NAPLS 3: $F_{2, 696} = 17.358$, p < 0.001), had higher general (Global Assessment of Functioning, GAF; NAPLS 2, $F_{4, 709} = 3.001$, $P = .018$, $\eta^2 = .017$; NAPL 3: $F_{4,689} = 4.906$, $P < 0.001$, $\eta^2 =. 028$) and social functioning (Global Functioning Social Scale (GFS), NAPLS 2: $F_{4, 705} = 4.102$, $P = 0.002$, $\eta^2 = .023$; NAPLS 3: $F_{4,692} = 5.987$, $p < 0.001$, $\eta^2 = 0.033$) and less severe depressive symptoms (NAPLS 2: $F_{4,673} = 5.686$, $P < .001$, $\eta^2 = 0.033$; NAPLS 3: $F_{4,692} = 3.957$, $P = 0.003$, $\eta^2 =. 023$). Rates of conversion to psychosis did not differ significantly based on symptom profile, but ORs trended toward a higher chance of conversion for participants with delusions as a first symptom compared with those reporting hallucinations (OR = 2.20) both symptoms (OR = 2.54) at onset (**Table S16 &S17**).

*Symptom transition probabilities*

To calculate probabilities of transition between states, the number of transitions between states was counted, and then the transition probability for each interval was calculated by dividing the count by the total count of a specific state; finally, the probabilities were averaged, and z-tests for differences between proportions were conducted (**Fig 2c,f,l; Table S18 to S20**).

*Supplementary Analysis: data from the Topography of Psychotic Episode (TOPE) in the PEPP sample*

Additional evidence of the relative stability of delusions to hallucinations can be found in the data from the Topography of Psychotic Episode (TOPE) assessment collected on a subset (n = 434) of the patients in the PEPP sample. In this assessment, the presence or absence of symptoms prior to the prodrome and their course was collected. With respect to course, symptoms were rated as not having been present during the prodrome; having been present during one phase or time period in the prodrome; having been present during several phases or time periods; having been present in a persistent manner during the prodrome; having onset only after psychosis onset; or representing a lifelong pattern of behavior. We note that (expected, given this is a retrospective data about the prodromal phase) a low number of delusion and hallucination symptoms should lead to cautious interpretations of findings from assessment of those symptoms.

Comparing Delusion spectrum symptoms (Odd/bizarre ideas, suspiciousness/ideas of reference, and delusions) to Hallucination spectrum samples (Unusual perceptual experiences, hallucinations), we see several important patterns (Table below). Firstly, it is more common for patients to experience delusion spectrum symptoms than hallucination spectrum symptoms, with more than double the number of patients reporting the most common delusion spectrum symptom (suspiciousness/ideas of reference, 39%) compared to the most common hallucination spectrum symptom (unusual perceptual experiences, 18%). When examining the course of symptoms and ignoring hallucinations and delusions (due to their low incidence), we see that the most persistent symptom is again suspiciousness/ideas of reference, with 21% of patients experiencing this symptom persistently during the prodrome, compared to 7.4% of patients experiencing perceptual alterations in a persistent manner. While this data is limited by its retrospective nature, these results support the finding that delusion-spectrum symptoms are both more common and more persistent in the prodrome.

Finally, while there is significant missing data in the TOPE for onset dates (~20%), we can examine whether patients tended to experience delusion and hallucination symptoms separately at onset, or at the same time. Of 45 patients for whom onset dates are available for both suspiciousness/ideas of reference and unusual perceptual experiences, 24 (53%) had the same onset date for both. In the 34 patients for whom onset dates were available for both odd/bizarre ideas and unusual perceptual experiences, 15 (44%) had the same onset date for both. In interpreting this one must be wary of recall bias; one might expect dates to be merged in recall when recall of specific symptom onsets were unclear. However, the conclusion can be drawn that these symptoms do not always begin at the same time, and from the results reported in the main paper, delusion-spectrum symptoms tend to be reported first, and they are reported more frequently overall, which again supports the primordial nature of hallucination-like symptoms. This also suggests that a larger proportion of those who had unusual perceptual experiences also had delusion-spectrum symptoms than vice versa: 45 patients have onset dates for both suspiciousness/ideas of reference and unusual perceptual experiences. Therefore at least 45 of 78 patients with unusual perceptual experiences (58%) also experienced suspiciousness; however, that also means at least 45 of 168 patients with suspiciousness (27%) also experienced unusual perceptual experiences. As such, it seems that it was more common to experience suspiciousness in isolation than in combination with perceptual experiences.

## Supplemental Tables

**Table S1. NAPLS 2 and 3 demographic information**

|         |            |     | Age   |      | Sex    | Race/ Ethnicity |       |       |       |
|---------|------------|-----|-------|------|--------|----------|-------|-------|-------|
|         |            | N   | Mean  | SD   | % Male | Hispanic | White | Black | Other |
| NAPLS 2 | Hall - Del | 80  | 17.31 | 3.82 | 50.00  | 21.30    | 61.30 | 18.80 | 19.90 |
|         | Del - Hall | 159 | 18.59 | 3.95 | 58.50  | 18.90    | 59.70 | 17.60 | 22.70 |
|         | Del & Hall | 275 | 18.61 | 4.39 | 54.20  | 17.50    | 59.30 | 4.50  | 36.20 |
|         | Hall       | 40  | 15.85 | 2.83 | 52.50  | 12.80    | 61.50 | 5.10  | 33.40 |
|         | Del        | 165 | 19.28 | 4.28 | 61.80  | 18.80    | 47.90 | 18.80 | 33.30 |
| NAPLS 3 | Hall - Del | 105 | 17.20 | 3.49 | 46.70  | 20.00    | 58.10 | 9.50  | 32.40 |
|         | Del - Hall | 154 | 17.92 | 4.07 | 51.90  | 20.80    | 48.70 | 13.00 | 38.30 |
|         | Del & Hall | 240 | 17.82 | 3.80 | 50.80  | 23.30    | 57.90 | 11.30 | 30.80 |
|         | Hall       | 21  | 16.52 | 3.82 | 61.90  | 14.30    | 61.90 | 14.30 | 23.80 |
|         | Del        | 179 | 19.63 | 4.25 | 63.10  | 16.80    | 56.20 | 11.20 | 32.60 |

Note. Age differences were found in NAPLS 2 and 3 (NAPLS 2, $F_{(4,714)} = 7.305$, $\eta^2 = .039$, multiple comparisons: Hall < Del-Hall, Del & Hall and Del, Del > Hall-Del, $Ps < 0.01$; NAPLS 3, $F_{(4,694)} = 9.273$, $\eta^2 = .051$, $Ps < 0.001$, multiple comparisons, Del > Del-Hall, Hall- Del, Del & Hall and Hall, $Ps < 0.01$). No association between sex and first symptom was found in NAPLS 2 ($\chi^2(4) = 4.379$, $P$ = ns). However, those who presented delusions at first symptom were more likely to be males (Del-Hall: $Z = 2.141$, $P = 0.032$, Del: $Z = 3.036$, $P = 0.002$). In NAPLS 3, we did find an association between sex and first symptom ($\chi^2(4)_{4MO} =10.033$, $P = 0.040$) among those whose first symptoms were delusions, males were more predominant ($Z = 3.513$, $P < 0.001$).

**Table S2. NAPLS 2. Frequency and percentage of symptoms reported from baseline to 24 months**

| Visit | | NS | Del | Hall | Del&Hall | Total | $\chi^2(1)$ |
|---|---|---|---|---|---|---|---|
| BL | Fr | 0 | 165 | 40 | 514 | 719 | 11.278* |
| | % | 0 | 22.9 | 5.6 | 71.5 | 100 | |
| M6 | Fr | 106 | 135 | 36 | 209 | 486 | 47.04* |
| | % | 21.6 | 28 | 7.4 | 43 | 100 | |
| M12 | Fr | 112 | 111 | 35 | 135 | 394 | 34.93* |
| | % | 28.5 | 28.2 | 8.9 | 34.4 | 100 | |
| M18 | Fr | 95 | 80 | 24 | 63 | 262 | 15.65* |
| | % | 36.3 | 30.5 | 9.1 | 24 | 100 | |
| M24 | Fr | 109 | 67 | 22 | 64 | 262 | 29.10* |
| | % | 41.6 | 25.6 | 8.4 | 24.4 | 100 | |

Note. NS = No symptoms; Del = Only Delusions; Hall = Only Hallucinations; Del&Hall = Delusions and Hallucinations; Fr = Frequency; BL = Baseline; M6 = 6 months; M12 = 12 months; M18 = 18 months; M24 = 24 months. *p < 0.001. Chi-square as a test of independence was conducted comparing delusions and hallucinations as dichotomous categories. The chi-square of goodness of fit test comparing 3 and 4 categories was significant at each time point $\chi^2(2)_{BL} = 256.185$, $\chi^2(3)_{6MO} = 71.867$, $\chi^2(3)_{12MO} = 26.159$, $\chi^2(3)_{18MO} = 37.489$, $\chi^2(3)_{24MO} = 32.504$, ps < 0.001).

**Table S3. NAPLS 2. Frequency of CHR symptoms at each timepoint in participants who converted to psychosis**

| | NS | Del | Hall | Del&Hall | Total |
|---|---|---|---|---|---|
| BL | 0 | 14 | 4 | 55 | 73 |
| M6 | 4 | 9 | 0 | 20 | 33 |
| M12 | 2 | 5 | 0 | 5 | 12 |
| M18 | 3 | 1 | 1 | 2 | 7 |
| M24 | 2 | 0 | 1 | 1 | 4 |

Note. 73 participants reached a severity score of 6 during the study. On average, participants who converted to psychosis did so 11.32 (SD = 9.27) months after baseline. Of those, 9.2% (6) experienced hallucinations, 47.7% (31) experienced delusions, and 43.1% (28) experienced both at the conversion visit. Participants who converted were 3.5 times more likely to present delusion only than hallucinations only at baseline (Z = 2.513, P = 0.011), and 13.75 times more likely to have hallucinations and delusions together than only hallucinations at baseline (Z = 8.6012, P < 0.001).

**Table S4. NAPLS 2. Frequency of CHR symptoms at each timepoint in participants who persisted**

|     | NS | Del | Hall | Del&Hall | Total |
|-----|----|-----|------|----------|-------|
| BL  | 0  | 67  | 13   | 244      | 324   |
| M6  | 23 | 89  | 22   | 159      | 293   |
| M12 | 18 | 82  | 21   | 116      | 237   |
| M18 | 15 | 68  | 18   | 57       | 158   |
| M24 | 0  | 67  | 21   | 63       | 151   |

Note. Throughout the study, symptoms persisted in 324 participants. The likelihood of having only delusions was 5.15 times greater than that of having only hallucinations (Z = 6.4486, $P < 0.001$). Moreover, this likelihood was markedly higher, at 18.76 times, when compared to the occurrence of both delusions and hallucinations together (Z =18.55, $P < 0.001$).

**Table S5. NAPLS 2. Frequency of CHR symptoms at each time point in participants who remitted**

|     | NS  | Del | Hall | Del&Hall | Total |
|-----|-----|-----|------|----------|-------|
| BL  | 0   | 44  | 17   | 110      | 171   |
| M6  | 77  | 38  | 14   | 30       | 159   |
| M12 | 92  | 24  | 14   | 14       | 144   |
| M18 | 77  | 11  | 5    | 4        | 97    |
| M24 | 107 | 0   | 0    | 0        | 107   |

Note. Throughout the study, symptoms persisted in 171 participants. The likelihood of having only delusions was 2.58 times greater than that of having only hallucinations (Z = 3.8138, $P < 0.001$). Moreover, this likelihood was markedly higher, at 6.47 times, when compared to the occurrence of both delusions and hallucinations together (Z = 10.4082, $P < 0.001$). In the NAPLS 2 study, 26.14% of them (188) were assessed only at the baseline stage. These participants were not included in the symptom persistence or remitter group.

**Table S6. NAPLS 3. Frequency and percentage of symptoms reported from baseline to 24 months**

| Visit | | NS | Del | Hall | Del&Hall | Total | $\chi^2(1)$ |
|---|---|---|---|---|---|---|---|
| BL | Fr | 0 | 179 | 21 | 499 | 699 | 7.453* |
| | % | 0 | 25.6 | 3.0 | 71.4 | 100 | |
| M2 | Fr | 38 | 130 | 30 | 285 | 483 | 14.47** |
| | % | 7.9 | 26.9 | 6.2 | 59 | 100 | |
| M4 | Fr | 69 | 125 | 25 | 216 | 435 | 38.80** |
| | % | 15.9 | 28.7 | 5.7 | 49.7 | 100 | |
| M6 | Fr | 81 | 102 | 34 | 178 | 395 | 36.56** |
| | % | 20.5 | 25.8 | 8.6 | 45.1 | 100 | |
| M8 | Fr | 98 | 111 | 39 | 149 | 397 | 28.79** |
| | % | 24.7 | 28 | 9.8 | 37.5 | 100 | |
| M12 | Fr | 112 | 112 | 30 | 112 | 366 | 29.31** |
| | % | 30.6 | 30.6 | 8.2 | 30.6 | 100 | |
| M18 | Fr | 96 | 79 | 26 | 87 | 288 | 27.23** |
| | % | 33.3 | 27.4 | 9 | 30.2 | 100 | |
| M24 | Fr | 99 | 72 | 21 | 53 | 245 | 16.84** |
| | % | 40.4 | 29.4 | 8.6 | 21.6 | 100 | |

Note. NS = No symptoms; Del = Only Delusions; Hal = Only Hallucinations; Del&Hall = Delusions and Hallucinations; BL = Baseline; M2 = 2 months; M4 = 4 months; M6 = 6 months; M8 = 8 months; M12 = 12 months; M18 = 18 months; M24 = 24 months. Chi-square as a test of independence was conducted, comparing delusions and hallucinations as dichotomous categories. The chi-square of goodness of fit test comparing 3 and 4 categories were significant at each time point $\chi^2(2)_{BL}$ =509.08, $\chi^2(3)_{2MO}$ = 349.04, $\chi^2(3)_{4MO}$ =187.23, $\chi^2(3)_{6MO}$ =109.35, $\chi^2(3)_{8MO}$ = 62.92, $\chi^2(3)_{12MO}$ = 55.11, $\chi^2(3)_{18MO}$ = 41.19, $\chi^2(3)_{24MO}$ = 52.71, $Ps < 0.001$).

**Table S7. NAPLS 3. Frequency of CHR symptoms at each time point in participants who converted to psychosis**

| | NS | Del | Hall | Del&Hall | Total |
|---|---|---|---|---|---|
| BL | 0 | 21 | 0 | 45 | 66 |
| M2 | 2 | 12 | 0 | 33 | 45 |
| M4 | 3 | 8 | 1 | 24 | 36 |
| M6 | 3 | 2 | 0 | 16 | 21 |
| M8 | 2 | 3 | 0 | 12 | 17 |
| M12 | 3 | 6 | 0 | 4 | 13 |
| M18 | 0 | 5 | 0 | 4 | 9 |
| M24 | 0 | 1 | 0 | 1 | 2 |

Note. 66 (9.4%) participants reached a severity score of 6 during the study. On average, participants who developed psychosis did so 10.27 (SD = 9.94) months after baseline. Of those, 8.2% (5) experienced hallucinations, 54.1% (33 experienced delusions, and 37.7% (23) experienced both at conversion visit. None of the participants reported having only hallucinations at baseline.

**Table S8. NAPLS 3. Frequency of CHR symptoms at each time point in participants who persisted**

|     | NS | Del | Hall | Del&Hall | Total |
|-----|----|-----|------|----------|-------|
| BL  | 0  | 74  | 13   | 278      | 365   |
| M2  | 8  | 71  | 19   | 195      | 293   |
| M4  | 16 | 76  | 16   | 157      | 265   |
| M6  | 21 | 63  | 24   | 133      | 241   |
| M8  | 19 | 77  | 24   | 117      | 237   |
| M12 | 20 | 87  | 22   | 98       | 227   |
| M18 | 9  | 64  | 20   | 77       | 170   |
| M24 | 0  | 71  | 21   | 52       | 144   |

Note. Throughout the study, symptoms persisted in 365 participants. The likelihood of having only delusions was 5.69 times greater than that of having only hallucinations ($Z = 6.9683$, $P < 0.001$). Moreover, this likelihood was markedly higher, at 21.38 times, when compared to the occurrence of both delusions and hallucinations together ($Z = 20.0322$, $P < 0.001$).

**Table S9. NAPLS 3. Frequency of CHR symptoms at each time point in participants who remitted**

|     | NS | Del | Hall | Del&Hall | Total |
|-----|----|-----|------|----------|-------|
| BL  | 0  | 57  | 8    | 114      | 179   |
| M2  | 8  | 71  | 19   | 195      | 203   |
| M4  | 50 | 41  | 8    | 35       | 134   |
| M6  | 57 | 37  | 10   | 29       | 133   |
| M8  | 77 | 31  | 15   | 20       | 143   |
| M12 | 89 | 19  | 8    | 10       | 126   |
| M18 | 87 | 10  | 6    | 6        | 109   |
| M24 | 99 | 0   | 0    | 0        | 99    |

Note. Throughout the study, symptoms persisted in 179 participants. The likelihood of having only delusions was 7.1 times greater than that of having only hallucinations ($Z = 6.7181$, $P < 0.001$). Moreover, this likelihood was markedly higher, at 14.25 times, when compared to the occurrence of both delusions and hallucinations together ($Z = 11.8198$, $P < 0.001$). In the NAPLS 3 study, 14.87% (104) were assessed only at the baseline stage. These participants were not included in the symptom persistence or remitter group.

**Table S10. First symptom onset for the Del&Hall group reported retrospectively at baseline**

|                                               | NAPLS 2 |        | NAPLS 3 |        |
|-----------------------------------------------|---------|--------|---------|--------|
| First Symptom                                 | Frq     | %      | Frq     | %      |
| Delusion before Hallucinations (Del-Hall)     | 161     | 43.8%  | 150     | 40.0%  |
| Hallucinations before Delusions (Hall-Del)    | 81      | 22.0%  | 104     | 27.3%  |
| Same onset for Delusions and Hallucinations (D&H) | 126 | 34.2%  | 121     | 32.2%  |
| Total                                         | 368     | 100.0% | 375     | 100.0% |

Note. First symptom onset date for Del&Hall group was explored in NAPLS 2 and 3. If the difference in the onset of symptoms was larger than a month, participants were grouped Del-Hall or Hall-Del. If the data was unavailable or they reported both symptoms at baseline, they were classified in Del&Hall group.

**Table S11. PEPP. Frequency and percentage of symptoms from first prodromal symptom to 24 months**

| Time | | NS/O | Del | Hall | Del&Hall | Total | $\chi^2(1)$ |
|---|---|---|---|---|---|---|---|
| FPr | Fr | 413 | 131 | 39 | NA | 583 | 194.466** |
| | % | 70.8 | 22.5 | 6.7 | NA | 100 | |
| FPs | Fr | 78 | 324 | 116 | NA | 518 | 92.987** |
| | % | 15.1 | 62.5 | 22.4 | NA | 100 | |
| FPs1 | Fr | 75 | 304 | 114 | NA | 493 | 84.098** |
| | % | 15.2 | 61.7 | 16.4 | NA | 100 | |
| BL | Fr | 12 | 187 | 7 | 471 | 677 | 9.129** |
| | % | 1.8 | 26.9 | 1 | 67.8 | 100 | |
| M1 | Fr | 132 | 116 | 21 | 109 | 378 | 47.126** |
| | % | 34.9 | 30.7 | 5.6 | 28.9 | 100 | |
| M2 | Fr | 231 | 128 | 27 | 92 | 478 | 60.746** |
| | % | 48.3 | 26.8 | 5.6 | 19.2 | 100 | |
| M3 | Fr | 319 | 125 | 28 | 107 | 563 | 110.478** |
| | % | 55.1 | 21.6 | 4.8 | 18.5 | 100 | |
| M6 | Fr | 293 | 119 | 30 | 94 | 536 | 87.634** |
| | % | 54.7 | 22.2 | 5.6 | 17.5 | 100 | |
| M9 | Fr | 262 | 92 | 28 | 76 | 458 | 76.740** |
| | % | 57.2 | 20.1 | 6.1 | 16.6 | 100 | |
| M12 | Fr | 270 | 94 | 27 | 83 | 474 | 86.819** |
| | % | 57.0 | 19.8 | 5.7 | 17.5 | 100 | |
| M18 | Fr | 246 | 78 | 27 | 78 | 429 | 84.244** |
| | % | 57.3 | 18.2 | 6.3 | 18.2 | 100 | |
| M24 | Fr | 241 | 87 | 18 | 65 | 411 | 74.024** |
| | % | 58.6 | 21.2 | 4.4 | 15.8 | 100 | |

Note. FPr = First prodromal symptom; FPs = First psychotic symptom; FPs1 = First psychotic symptom lasting at least one week; NS/O = No symptoms or "other" prior to baseline; Del = Delusions only; Hall = Hallucinations only; Del&Hall = Delusions and hallucinations; NA = Not applicable. FPr, FPs, and FPs1 were retrospectively collected.

**Table S12. NAPLS 2 functioning and cognitive measures at baseline**

|  |  | GAF | GFS | GFR | CDS | DSI Frq | DSI Total | MAP Task | WRAT | WASI | BACS | HLVT |
|---|---|---|---|---|---|---|---|---|---|---|---|---|
| Hall - Del | N | 78 | 78 | 78 | 75 | 72 | 70 | 65 | 75 | 75 | 75 | 75 |
|  | M | 49.14 | 6.26 | 5.78 | 6.56 | 26.44 | 91.51 | 515.84 | 101.04 | 101 | 54.96 | 24.96 |
|  | SD | 9.03 | 1.52 | 2.02 | 5.26 | 23.61 | 73.7 | 164.46 | 16.89 | 15.44 | 13.03 | 5.33 |
| Del - Hall | N | 159 | 158 | 157 | 152 | 144 | 140 | 128 | 140 | 138 | 137 | 138 |
|  | M | 47.7 | 6.23 | 5.91 | 6.72 | 27.64 | 112.26 | 544.8 | 106.25 | 104.58 | 57.45 | 26.25 |
|  | SD | 10.42 | 1.6 | 2.27 | 4.63 | 24.91 | 153.79 | 276.61 | 16.16 | 15.45 | 13.55 | 4.67 |
| Del & Hall | N | 273 | 270 | 269 | 258 | 243 | 238 | 226 | 245 | 245 | 244 | 245 |
|  | M | 49.17 | 6.32 | 6.04 | 5.66 | 26.14 | 99.87 | 502.94 | 103.93 | 103.84 | 57.45 | 26.2 |
|  | SD | 11.08 | 1.51 | 2.24 | 4.59 | 20.74 | 88.58 | 210.9 | 15.03 | 14.39 | 12.71 | 5.01 |
| Del | N | 164 | 164 | 164 | 154 | 145 | 143 | 125 | 149 | 148 | 150 | 150 |
|  | M | 47.49 | 5.8 | 5.95 | 5.55 | 28.23 | 102.97 | 552.97 | 107.82 | 104.83 | 57.1 | 24.98 |
|  | SD | 10.78 | 1.59 | 2.11 | 4.65 | 27.02 | 128.4 | 210.51 | 16.97 | 15.36 | 11.25 | 4.98 |
| Hall | N | 40 | 40 | 40 | 39 | 35 | 34 | 32 | 37 | 36 | 37 | 37 |
|  | M | 53.48 | 6.73 | 6.15 | 2.97 | 19.46 | 62.29 | 516.75 | 107.84 | 102.58 | 55.08 | 24.65 |
|  | SD | 12.47 | 1.68 | 1.89 | 4.46 | 9.48 | 41.3 | 199.52 | 19.5 | 18.11 | 12.62 | 5.32 |

Note. GAF = Global Assessment of Functioning, measures how much a person's symptoms affect their day-to-day life on a scale of 0 to 100, those with delusion only at baseline showed lower functioning than Hall and Del-Hall ($F_{4,705}$ = 3.001, $P$ = 0.018, $\eta^2$ = .017, Del < Hall > Hall-Del); GFS = Global Functioning Social Scale rated from 1 to 10, with high scores indicating better functioning, Del only at baseline showed lower scores

than Hall and Del&Hall ($F_{4,705}$ = 4.202, $P$ = 0.002, $\eta^2$ =. 023, Del < Hall & Del-Hall, $Ps$ < 0.05); GFR = Global Functioning Role scale, rated from 1 to 10, with high scores indicating better functioning ($\eta^2$ = .002, $Ps$ = ns); CDS = Calgary Depression Scale ($F_{4,673}$ = 5.686, $P$ < 0.001, $\eta^2$ = .033, H < all groups, $Ps$ < 0.05); DSI Frq = Daily Stress Inventory frequency of events ($\eta^2$ = .007, $P$ = ns); DSI total = Daily Stress Inventory average of distress ($\eta^2$ = .009, $P$ = ns); MAP Task = Functional Capacity Map Task (Blocks) ($\eta^2$ = .005, $P$ = ns); WRAT reading = Wide Range Achievement Test reading standard scores ($F_{4,641}$ = 2.967, $P$ = 0.019, no significant pairwise comparisons); WASI = Wechsler Abbreviated Scale of Intelligence ($P$ = ns); BACS = Brief Assessment of Cognition in Schizophrenia, Simbol Coding ($\eta^2$ = .005, $P$ = ns); HLVT = Hopkins Verbal Learning Test–Revised ($F_{4,640}$ = 2.644, $P$ = 0.033, $\eta^2$ =. 016, no significant pairwise comparisons $Ps$ = ns). Participants who transitioned to psychosis performed similarly on functioning (GAF, GFS, GFR, MAP Task), level of depression (CDS), frequency of distress events (DSI Frq), and distress(DSI) compared with those who didn't convert. However, converters did poorly on BACS ($T_{(641)}$ = 2.795, $P$ = 0.003) and HLVT ($T_{(643)}$ = 2.524, $P$ = 0.006).

**Table S13. NAPLS 3 functioning and cognitive measures at baseline**

| | | GAF | GFS | GFR | CDS | DSI Frq | DSI Total | Go to Town | WRAT | WASI | BACS | HLVT |
|---|---|---|---|---|---|---|---|---|---|---|---|---|
| Hall - Del | N | 104 | 104 | 57 | 103 | 100 | 97 | 99 | 100 | 101 | 101 | 101 |
| | M | 52.66 | 6.81 | 1.47 | 6.11 | 21.72 | 76.53 | 3.82 | 107.01 | 102.54 | 52.48 | 26.11 |
| | SD | 12.737 | 1.401 | 0.889 | 4.43 | 12.129 | 54.334 | 2.915 | 17.133 | 16.646 | 13.812 | 5.51 |
| Del - Hall | N | 152 | 154 | 86 | 152 | 149 | 148 | 137 | 147 | 147 | 148 | 148 |
| | M | 49.48 | 6.35 | 1.49 | 6.99 | 21.23 | 75.26 | 3.53 | 111.57 | 106.05 | 56.07 | 26.82 |
| | SD | 12.004 | 1.462 | 0.837 | 4.565 | 11.597 | 57.259 | 2.865 | 18.884 | 15.819 | 13.246 | 5.477 |
| Del & Hall | N | 238 | 239 | 135 | 233 | 223 | 223 | 216 | 230 | 228 | 228 | 227 |
| | M | 50.39 | 6.43 | 1.39 | 6.67 | 21.85 | 75.1 | 3.56 | 108.67 | 106.52 | 55.84 | 26.44 |
| | SD | 11.876 | 1.487 | 0.81 | 4.495 | 11.335 | 56.748 | 2.907 | 17.233 | 14.971 | 13.063 | 5.035 |
| Del | N | 179 | 179 | 96 | 178 | 172 | 171 | 168 | 176 | 176 | 177 | 177 |
| | M | 50.75 | 6.14 | 1.61 | 5.81 | 18.83 | 59.23 | 3.67 | 110.37 | 105.62 | 52.42 | 26.09 |
| | SD | 10.974 | 1.596 | 0.91 | 4.191 | 10.117 | 43.43 | 3.259 | 16.319 | 17.183 | 13.576 | 5.056 |
| Hall | N | 21 | 21 | 1 | 21 | 20 | 19 | 21 | 21 | 21 | 21 | 21 |
| | M | 60.81 | 7.48 | 1 | 3.57 | 13.4 | 29.68 | 4.43 | 109.43 | 104.38 | 54.57 | 26.67 |
| | SD | 13.129 | 1.537 | - | 3.429 | 7.701 | 22.413 | 2.675 | 15.5 | 13.204 | 9.347 | 4.127 |

Note. GAF = Global Assessment of Functioning, ($F_{4,689}$ = 4.906, $p < .001$, $\eta^2 = .028$, Hall > all groups, $P$s <.05); GFS = Global Functioning Social scale, ($F_{4,692}$ = 5.987, $P < .001$, $\eta^2 = .033$, Hall > Del-Hall, Del&Hall and Del; Del < Hall-Del and Hall, $P$s <.05); GFR = Global

Functioning Role scale; GFR = Global Functioning Role scale ($\eta^2$ = .012); CDS = Calgary Depression Scale, ($F_{4,692}$ = 3.957, $P$ = .003, $\eta^2$ = .023; Hall < Del-Hall and Del&Hall, $P$s < .05); DSI Frq = Daily Stress Inventory frequency of events ($F_{4,659}$ =4.257, $P$ < .001, $\eta^2$ = .025; Hall < all groups except Del, $P$s <.05); DSI total = Daily Stress Inventory average of distress ($F_{4,659}$ =4.257, $P$ = .002, $\eta^2$ = .035; Hall < all groups except Del, $P$s <.05); Go to Town = Functional Capacity (errands completed) ($P$ = ns, $\eta^2$ = .003); WRAT reading = Wide Range Achievement Test reading standard scores ($F_{4,636}$ = 4.684, $P$ = .001, $\eta^2$ = .008; no significant differences after multiple comparisons); WASI = Wechsler Abbreviated Scale of Intelligence ($P$ = ns, $\eta^2$ = .007); BACS = Brief Assessment of Cognition in Schizophrenia, Simbol Coding, $F_{(4,670)}$ = 2.779, $P$ = .026, $\eta^2$ = .016, do not suvided multiple comparisons; HLVT = Hopkins Verbal Learning Test–Revised ($P$ = ns, $\eta^2$ = .003). Participants who transitioned to psychosis performed poorly in cognitive and daily functioning compared with those who didn't convert ($T_{GAF (692)}$ = 3.995, $P$ < 0.000; $T_{GFS (695)}$ = 3.091, $P$ = 0.002; $T_{BACS (673)}$ = 2.772, $P$ = 0.006; $T_{HLVT (672)}$ = 2.099, $P$ = 0.036; $T_{WASI (671)}$ = 2.213, $P$ = 0.027), but they have the same levels of depression (CDS), frequency of daily stress events (DSI Frq) and daily stress events (DSI), same performance on WRAT and the Go to Town Task.

**Table S14. Prevalence in percentages of DSM-IV diagnoses at baseline in NAPLS 2 and NAPLS 3 samples**

|  | NAPLS 2 | | | | | NAPLS 3 | | | | |
| --- | --- | --- | --- | --- | --- | --- | --- | --- | --- | --- |
|  | Hall - Del (n = 79) | Del - Hall (n = 156) | Del&Hall (n = 265) | Hall (n = 39) | Del (n = 160) | Hall - Del (n = 104) | Del - Hall (n = 153) | Del&Hall (n = 236) | Hall (n = 21) | Del (n = 179) |
| Depression | 50.63 | 53.21 | 36.98 | 28.21 | 44.37 | 50 | 53.59 | 49.58 | 23.81 | 44.13 |
| Bipolar | 3.8 | 9.62 | 6.04 | 0 | 9.37 | 6.73 | 9.87 | 5.51 | 0 | 6.15 |
| OCD | 7.59 | 7.05 | 6.79 | 2.56 | 7.5 | 12.5 | 9.15 | 12.71 | 0 | 9.5 |
| PTSD | 2.53 | 2.56 | 3.4 | 2.56 | 0 | 10.58 | 5.88 | 8.9 | 0 | 7.26 |
| Panic Disorder | 12.66 | 14.1 | 11.32 | 0 | 10 | 24.04 | 17.65 | 23.73 | 14.29 | 11.17 |
| Agoraphobia | 1.27 | 0.64 | 1.89 | 0 | 0 | 8.65 | 7.84 | 9.75 | 0 | 3.35 |
| Social Phobia | 11.39 | 19.23 | 13.21 | 10.26 | 14.37 | 22.12 | 23.53 | 22.88 | 4.76 | 22.91 |
| Specific Phobias | 10.13 | 13.46 | 10.19 | 17.95 | 5.62 | 10.58 | 7.19 | 10.17 | 9.52 | 5.59 |
| GAD | 26.58 | 23.08 | 20.75 | 17.95 | 28.12 | 30.77 | 28.76 | 29.24 | 23.81 | 34.64 |
| ADHD | 18.99 | 18.59 | 15.09 | 28.95 | 18.75 | 22.12 | 18.3 | 16.95 | 23.81 | 17.32 |
| BPD | 9.33 | 3.4 | 4.37 | 0 | 1.91 | 1.92 | 1.96 | 6.36 | 4.76 | 5.59 |
| SPD | 16.46 | 14.74 | 21.51 | 5.13 | 19.88 | 8.65 | 10.46 | 13.98 | 0 | 11.73 |

Note. DSM-IV diagnoses based on the Structured Clinical Interview for DSM-IV (SCID-IV); Bipolar = Bipolar Disorder; OCD = Obsessive-Compulsive Disorder; PTSD = Post-Traumatic Stress Disorder; GAD = Generalized Anxiety Disorder; ADHD = Attention-Deficit/ Hyperactivity Disorder; BPD = Borderline Personality Disorder; SPD = Schizotypal Personality Disorder.

**Table S15. Prevalence of diagnoses at baseline in the PEPP sample of first reported prodromal, first reported psychotic symptoms and based on symptoms present at baseline**

|  | First Prodromal Symptom | | First Psychotic Symptom | | Baseline (study start) | | |
| --- | --- | --- | --- | --- | --- | --- | --- |
|  | Hall only (n = 39) | Del only (n = 131) | Hall only (n = 116) | Del only (n = 324) | Hall & del (n = 471) | Hall only (n = 7) | Del only (n = 187) |
| SOFAS | M: 41.08 (13.383) | M: 43.34 (12.790) | M: 42.50 (11.797) | M: 40.14 (12.066) | M: 40.49 (13.053) | M: 53.86 (16.945) | M: 41.09 (12.692) |
| Schizophrenia | 17 (43.59%) | 61 (46.56%) | 46 (39.66%) | 119 (36.73%) | 176 (37.37%) | 2 (28.57%) | 61 (32.62%) |
| Psychosis NOS | 10 (25.64%) | 16 (12.21%) | 20 (17.24%) | 54 (16.67%) | 72 (15.29%) | 2 (28.57%) | 25 (13.37%) |
| Schizoaffective disorder | 3 (7.69%) | 7 (5.34%) | 9 (7.76%) | 19 (5.86%) | 39 (8.28%) | 0 (0%) | 8 (4.28%) |
| Schizophreniform disorder | 3 (7.69%) | 8 (6.11%) | 4 (3.45%) | 10 (3.09%) | 20 (4.25%) | 0 (0%) | 3 (1.60%) |
| Brief psychotic disorder | 0 (0%) | 0 (0%) | 0 (0%) | 1 (0.31%) | 1 (0.21%) | 1 (14.29%) | 0 (0%) |
| Delusional disorder | 0 (0%) | 8 (6.11%) | 0 (0%) | 7 (2.16%) | 2 (0.42%) | 0 (0%) | 9 (4.81%) |
| Substance-induced psychosis | 1 (2.56%) | 2 (1.53%) | 0 (0%) | 2 (0.62%) | 1 (0.21%) | 1 (14.29%) | 2 (1.07%) |
| Bipolar Disorder | 2 (5.13%) | 20 (15.27%) | 14 (12.07%) | 60 (18.52%) | 66 (14.01%) | 1 (14.29%) | 36 (19.25%) |
| Depression* | 10 (25.64%) | 21 (16.03%) | 19 (16.38%) | 54 (16.67%) | 75 (15.92%) | 2 (28.57%) | 19 (10.16%) |
| Major depression with psychotic features | 3 (7.69%) | 10 (7.63%) | 21 (18.10%) | 49 (15.12%) | 55 (11.68%) | 0 (0%) | 22 (11.76%) |
| Mood Disorder | 0 (0%) | 1 (0.76%) | 0 (0%) | 1 (0.31%) | 1 (0.21%) | 0 (0%) | 0 (0%) |
| NOS | | | | | | | |
| Substance-induced mood disorder | 0 (0%) | 0 (0%) | 1 (0.86%) | 0 (0%) | 1 (0.21%) | 0 (0%) | 0 (0%) |
| Anxiety disorder | 0 (0%) | 14 (10.69%) | 10 (8.62%) | 28 (8.64%) | 35 (7.43%) | 0 (0%) | 12 (6.42%) |
| Panic disorder | 1 (2.56%) | 4 (3.05%) | 3 (2.59%) | 11 (3.40%) | 14 (2.97%) | 0 (0%) | 2 (1.07%) |
| OCD | 1 (2.56%) | 2 (1.53%) | 5 (4.31%) | 2 (0.62%) | 7 (1.49%) | 0 (0%) | 2 (1.07%) |
| PTSD | 1 (2.56%) | 1 (0.76%) | 5 (4.31%) | 4 (1.23%) | 7 (1.49%) | 0 (0%) | 3 (1.60%) |
| Alcohol abuse | 8 (20.51%) | 22 (16.79%) | 22 (18.97%) | 50 (15.43%) | 69 (14.65%) | 2 (28.57%) | 25 (13.37%) |
| Substance abuse | 21 (53.85%) | 84 (64.12%) | 61 (52.59%) | 196 (60.49%) | 249 (52.87%) | 3 (42.86%) | 99 (52.94%) |
| Hallucinogen dependence | 0 (0%) | 0 (0%) | 0 (0%) | 1 (0.31%) | 1 (0.21%) | 0 (0%) | 1 (0.53%) |
| Eating disorder | 0 (0%) | 0 (0%) | 1 (0.86%) | 0 (0%) | 1 (0.21%) | 0 (0%) | 0 (0%) |
| ADHD | 0 (0%) | 1 (0.76%) | 1 (0.86%) | 1 (0.31%) | 2 (0.42%) | 0 (0%) | 0 (0%) |
| Agoraphobia | 0 (0%) | 0 (0%) | 0 (0%) | 3 (0.93%) | 2 (0.42%) | 0 (0%) | 2 (1.07%) |
| Hypochondriasis | 1 (2.56%) | 0 (0%) | 1 (0.86%) | 0 (0%) | 1 (0.21%) | 0 (0%) | 0 (0%) |
| Social phobia | 2 (5.13%) | 6 (4.58%) | 8 (6.90%) | 13 (4.01%) | 17 (3.61%) | 0 (0%) | 5 (2.14%) |
| Specific phobia | 0 (0%) | 0 (0%) | 0 (0%) | 1 (0.31%) | 2 (0.42%) | 0 (0%) | 0 (0%) |

* Includes Dysthymia

**Table S16. NAPLS 2, odds ratios for conversion by symptoms at baseline and symptom onset.**

| | Baseline | | | | First Symptom | | |
|---|---|---|---|---|---|---|---|
| Symptom | Hall | Del | Del&Hall | Symptom | Hall | Del | Del&Hall |
| Hall (n = 4) | 1 | | | Hall (n = 9) | 1 | | |
| Del (n = 14) | 1.20 | 1 | | Del (n = 35) | 1.49 | 1 | |
| Del&Hall (n = 55) | 1.08 | 0.77 | 1 | Del&Hall (n = 29) | 1.45 | 0.98 | 1 |

**Note.** The table on the left shows the odds ratios of the symptoms at baseline, while the table on the right shows the odds ratios of first reported symptoms retrospectively. At baseline, individuals with only delusions had a slightly higher chance of converting than those with only hallucinations. The same trend was observed between delusions and hallucinations at baseline versus only hallucinations. Participants with only delusions had slightly fewer chances of conversion than those with both symptoms. No statistical differences were found in the proportions of conversion between the different baseline symptom presentations (Del vs Hall: Z = 0.3038, Hall vs Del&Hall; Z = 0.1383, Del vs Del&Hall: Z = 0.8195, $Ps$ = ns). The same trend was observed when the conversion rates for the first reported retrospective symptom were analyzed (Del first vs Hall first Z =1.0343, Hall first vs Del&Hall; Z = 1.0343; Del first vs Del&Hall Z = 0.1015, $Ps$ = ns).

**Table S17. NAPLS 3, odds ratios for conversion by symptom onset.**

| | Baseline | | | | First Symptom | | |
|---|---|---|---|---|---|---|---|
| Symptom | Hall | Del | Del&Hall | Symptom | Hall | Del | Del&Hall |
| Hall (n = 0) | 1 | | | Hall (n = 9) | 1 | | |
| Del (n = 21) | - | 1 | | Del (n = 35) | 2.20 | 1 | |
| Del&Hall (n = 45) | - | 1.34 | 1 | Del&Hall (n = 29) | 2.54 | 0.87 | 1 |

The table on the left displays the odds ratios of symptoms at the baseline, while the table on the right shows the odds ratios of first reported symptoms retrospectively. The same trend as NAPLS 2 was observed in the NAPLS 3 sample. However, participants who presented only hallucinations at baseline did not experience any conversions. The odds ratios considering the first symptom reported retrospectively indicated that the conversion rates were double for those who had delusions first as compared to those who had only hallucinations first. The same trend was observed for participants who reported retrospectively starting with both symptoms as compared to those who had only hallucinations. Finally, the ratio of conversion between delusions and both symptoms reported retrospectively was slightly lower for delusion only. No statistical differences were found in the conversion proportions between the different baseline symptom presentations (Del vs Del&Hall: $Z = 1.0508$, $P$ = ns). The same trend was observed when the conversion rates for the first reported retrospective symptom were analyzed (Del vs Del&Hall: $Z = 1.0508$, $P$ = ns, and as a first symptom: Del first vs Hall first $Z = 1.3162$, Hall first vs Del&Hall; $Z = 1.6086$; Del first vs Del&Hall $Z = 0.5169$, $Ps$ = ns).

**Table S18. NAPLS 2, symptom transition probabilities from visit to visit**

| Visit | Symt | NS Fr | NS % | D Fr | D % | H Fr | H % | DH Fr | DH % | Total |
|---|---|---|---|---|---|---|---|---|---|---|
| BL to M6 | NS | 0 | 0 | 0 | 0 | 0 | 0 | 0 | 0 | 0 |
|  | D | 29 | 26.4% | 68 | **61.8%** | 2 | 1.8% | 11 | 10.0% | 110 |
|  | H | 16 | 48.5% | 2 | 6.1% | 8 | **24.2%** | 7 | 21.2% | 33 |
|  | DH | 59 | 17.2% | 67 | 19.5% | 26 | 7.6% | 191 | **55.7%** | 343 |
|  | Total | 105 |  | 136 |  | 36 |  | 209 |  | 486 |
| M6 to M12 | NS | 58 | **72.5%** | 13 | 16.3% | 7 | 8.8% | 2 | 2.5% | 80 |
|  | D | 23 | 24.7% | 50 | **53.8%** | 3 | 3.2% | 17 | 18.3% | 93 |
|  | H | 10 | 35.7% | 2 | 7.1% | 13 | **46.4%** | 3 | 10.7% | 28 |
|  | DH | 14 | 9.2% | 33 | 21.7% | 12 | 7.9% | 93 | **61.2%** | 152 |
|  | Total | 105 |  | 98 |  | 35 |  | 115 |  | 353 |
| M12 to M18 | NS | 53 | **82.8%** | 8 | 12.5% | 1 | 1.6% | 2 | 3.1% | 64 |
|  | D | 20 | 29.4% | 36 | **52.9%** | 1 | 1.5% | 11 | 16.2% | 68 |
|  | H | 11 | 44.0% | 3 | 12.0% | 8 | **32.0%** | 3 | 12.0% | 25 |
|  | DH | 10 | 11.5% | 27 | 31.0% | 9 | 10.3% | 41 | **47.1%** | 87 |
|  | Total | 94 |  | 74 |  | 19 |  | 57 |  | 244 |
| M18 to M24 | NS | 65 | **81.3%** | 7 | 8.8% | 6 | 7.5% | 2 | 2.5% | 80 |
|  | D | 11 | 18.0% | 36 | **59.0%** | 2 | 3.3% | 12 | 19.7% | 61 |
|  | H | 5 | 27.8% | 1 | 5.6% | 10 | **55.6%** | 2 | 11.1% | 18 |
|  | DH | 4 | 8.5% | 7 | 14.9% | 3 | 6.4% | 33 | **70.2%** | 47 |
|  | Total | 85 |  | 51 |  | 21 |  | 49 |  | 206 |

Note. NS = No symptoms; D = Only delusions; H = Only hallucinations; DH = Delusions and hallucinations; Symt = Symptom; Fr = Frequency; BL = Baseline; M6 = 6 months; M12 = 12 months; M18 = 18 months; M24 = 24 months. Bold values indicate the probabilities of remaining with the same symptom from one time point to the next one (e.g., 61.8% of cases who reported only delusions at baseline persist with the same symptom at the 6-month follow-up).

**Table S19. NAPLS 3 symptom transitions probabilities from visit to visit**

| Visit | Symt | NS Fr | NS % | D Fr | D % | H Fr | H % | DH Fr | DH % | Total |
|---|---|---|---|---|---|---|---|---|---|---|
| BL to M2 | NS | 0 | 0 | 0 | 0 | 0 | 0 | 0 | 0 | 0 |
| | D | 18 | 14.6% | 84 | **68.3%** | 1 | 0.8% | 20 | 16.3% | 123 |
| | H | 3 | 15.8% | 0 | 0.0% | 14 | **73.7%** | 2 | 10.5% | 19 |
| | DH | 17 | 5.0% | 46 | 13.5% | 15 | 4.4% | 263 | **77.1%** | 341 |
| | Total | 38 | | 130 | | 30 | | 285 | | 483 |
| M2 to M4 | NS | 26 | **96.3%** | 1 | 3.7% | 0 | 0.0% | 0 | 0.0% | 27 |
| | D | 15 | 17.0% | 65 | **73.9%** | 0 | 0.0% | 8 | 9.1% | 88 |
| | H | 7 | 28.0% | 0 | 0.0% | 14 | **56.0%** | 4 | 16.0% | 25 |
| | DH | 14 | 6.5% | 36 | 16.8% | 9 | 4.2% | 155 | **72.4%** | 214 |
| | Total | 62 | | 102 | | 23 | | 167 | | 354 |
| M4 to M6 | NS | 45 | **86.5%** | 3 | 5.8% | 1 | 1.9% | 3 | 5.8% | 52 |
| | D | 16 | 18.0% | 56 | **62.9%** | 0 | 0.0% | 17 | 19.1% | 89 |
| | H | 2 | 9.5% | 1 | 4.8% | 15 | **71.4%** | 3 | 14.3% | 21 |
| | DH | 5 | 3.3% | 17 | 11.2% | 11 | 7.2% | 119 | **78.3%** | 152 |
| | Total | 68 | | 77 | | 27 | | 142 | | 314 |
| M6 to M8 | NS | 50 | **70.4%** | 9 | 12.7% | 9 | 12.7% | 3 | 4.2% | 71 |
| | D | 18 | 23.1% | 53 | **67.9%** | 0 | 0.0% | 7 | 9.0% | 78 |
| | H | 6 | 21.4% | 2 | 7.1% | 16 | **57.1%** | 4 | 14.3% | 28 |
| | DH | 5 | 3.5% | 22 | 15.5% | 11 | 7.7% | 104 | **73.2%** | 142 |
| | Total | 79 | | 86 | | 36 | | 118 | | 319 |
| M8 to M12 | NS | 68 | **86.1%** | 6 | 7.6% | 3 | 3.8% | 2 | 2.5% | 79 |
| | D | 14 | 16.1% | 60 | **69.0%** | 0 | 0.0% | 13 | 14.9% | 87 |
| | H | 8 | 24.2% | 2 | 6.1% | 16 | **48.5%** | 7 | 21.2% | 33 |
| | DH | 9 | 7.8% | 27 | 23.3% | 7 | 6.0% | 73 | **62.9%** | 116 |
| | Total | 99 | | 95 | | 26 | | 95 | | 315 |
| M12 to M18 | NS | 56 | **74.7%** | 14 | 18.7% | 4 | 5.3% | 1 | 1.3% | 75 |
| | D | 18 | 23.1% | 46 | **59.0%** | 2 | 2.6% | 12 | 15.4% | 78 |
| | H | 2 | 8.7% | 2 | 8.7% | 10 | **43.5%** | 9 | 39.1% | 23 |
| | DH | 6 | 7.5% | 13 | 16.3% | 8 | 10.0% | 53 | **66.3%** | 80 |
| | Total | 82 | | 75 | | 24 | | 75 | | 256 |
| M18 to M24 | NS | 60 | **87.0%** | 7 | 10.1% | 1 | 1.4% | 1 | 1.4% | 69 |
| | D | 10 | 18.2% | 38 | **69.1%** | 1 | 1.8% | 6 | 10.9% | 55 |
| | H | 6 | 33.3% | 1 | 5.6% | 8 | **44.4%** | 3 | 16.7% | 18 |
| | DH | 6 | 9.8% | 12 | 19.7% | 8 | 13.1% | 35 | **57.4%** | 61 |
| | Total | 82 | | 58 | | 18 | | 45 | | 203 |

Note. NS = No symptoms; Del = Only Delusions; Hall = Only Hallucinations; Del&Hall = Delusions and Hallucinations; Symt = Symptom; Fr = Frequency; BL = Baseline; M2 = 2 months; M4 = 4 months; M6 = 6 months; M8 = 8 months; M12 = 12 months; M18 = 18 months; M24 = 24 months. Bold values indicate the probabilities of remaining with the same symptom from one time point to the next one (e.g., 68.3% of cases who reported only delusions at baseline persist with the same symptom at the 6-month follow-up).

**Table S20. PEPP symptom probability transitions from visit to visit**

| Visit | Symt | NS/O Fr | % | Del Fr | % | Hall Fr | % | Del&Hall Fr | % | Total |
|---|---|---|---|---|---|---|---|---|---|---|
| FPr to FPs | O | 65 | 18.6% | 209 | 59.9% | 75 | 21.4% | 0 | NA | 349 |
| | D | 6 | 5.9% | 86 | **85.1%** | 9 | 8.9% | 0 | NA | 101 |
| | H | 2 | 6% | 7 | 21.2% | 24 | **72.7%** | 0 | NA | 33 |
| | DH | 0 | NA | 0 | NA | 0 | NA | 0 | NA | 0 |
| | Total | 73 | | 302 | | 108 | | 0 | | 483 |
| FPs to FPs1 | O | 72 | **96%** | 2 | 2.7% | 1 | 1.3% | 0 | NA | 75 |
| | D | 3 | 0.9% | 301 | **97.7%** | 4 | 1.3% | 0 | NA | 308 |
| | H | 0 | 0% | 1 | 0.9% | 109 | **99.1%** | 0 | NA | 110 |
| | DH | 0 | NA | 0 | NA | 0 | NA | 0 | NA | 0 |
| | Total | 75 | | | | | | 0 | | 493 |
| FPs1 to BL | O | 5 | **6.8%** | 25 | **34.2%** | 2 | 2.7% | 41 | 56.2% | 73 |
| | D | 0 | 0% | 101 | 33.9% | 0 | 0% | 197 | 66.1% | 298 |
| | H | 1 | 0.8% | 5 | 4.5% | 0 | 0% | 106 | **94.6%** | 112 |
| | DH | 0 | NA | 0 | NA | 0 | NA | 0 | NA | 0 |
| | Total | 6 | | 131 | | 2 | | 344 | | 483 |
| BL to M1 | NS | 3 | 50% | 2 | 33.3% | 1 | 16.7% | 0 | 0% | 6 |
| | D | 39 | 37.5% | 61 | **58.7%** | 0 | 0% | 4 | 3.8% | 104 |
| | H | 2 | **66.7%** | 0 | 0% | 1 | **33.3%** | 0 | **42.8%** | 3 |
| | DH | 66 | 26.4% | 58 | 23.2% | 19 | 7.6% | 107 | 30.6% | 250 |
| | Total | 110 | | 121 | | 21 | | 111 | | 363 |
| M1 to M2 | NS | 100 | **87.0%** | 12 | 10.4% | 2 | 1.7% | 1 | 0.9% | 115 |
| | D | 40 | 39.2% | 56 | **54.9%** | 3 | 2.9% | 3 | 2.9% | 102 |
| | H | 11 | 61.1% | 1 | 5.6% | 2 | **11.1%** | 4 | 22.2% | 18 |
| | DH | 19 | 19.4% | 20 | 20.4% | 9 | 9.2% | 50 | **51.0%** | 98 |
| | Total | 170 | | 89 | | 16 | | 58 | | 333 |
| M2 to M3 | NS | 192 | **86.5%** | 17 | 7.7% | 5 | 2.3% | 8 | 3.6% | 222 |
| | D | 41 | 33.9% | 68 | **56.2%** | 1 | 0.8% | 11 | 9.1% | 121 |
| | H | 14 | 53.8% | 0 | 0.0% | 6 | **23.1%** | 6 | 23.1% | 26 |
| | DH | 20 | 23.3% | 11 | 12.8% | 8 | 9.3% | 47 | **54.7%** | 86 |
| | Total | 267 | | 96 | | 20 | | 72 | | 455 |
| M3 to M6 | NS | 218 | **79.3%** | 31 | 11.3% | 7 | 2.5% | 19 | 6.9% | 275 |
| | D | 36 | 33.0% | 60 | **55.0%** | 1 | 0.9% | 12 | 11.0% | 109 |
| | H | 8 | 29.6% | 2 | 7.4% | 8 | **29.6%** | 9 | 33.3% | 27 |
| | DH | 19 | 19.6% | 19 | 19.6% | 14 | 14.4% | 45 | **46.4%** | 97 |
| | Total | 281 | | 112 | | 30 | | 85 | | 508 |
| M6 to M9 | NS | 190 | **80.5%** | 30 | 12.7% | 6 | 2.5% | 10 | 4.2% | 236 |
| | D | 35 | 37.6% | 45 | **48.4%** | 4 | 4.3% | 9 | 9.7% | 93 |
| | H | 9 | 42.9% | 0 | 0.0% | 4 | **19.0%** | 8 | 38.1% | 21 |
| | DH | 11 | 16.7% | 6 | 9.1% | 11 | 16.7% | 38 | **57.6%** | 66 |
| | Total | 245 | | 81 | | 25 | | 65 | | 416 |
| M9 to M12 | NS | 175 | **76.4%** | 30 | 13.1% | 6 | 2.6% | 18 | 7.9% | 229 |

|  |  |  |  |  |  |  |  |  |  |
|---|---|---|---|---|---|---|---|---|---|
|  | D | 34 | 41.0% | 38 | **45.8%** | 1 | 1.2% | 10 | 12.0% | 83 |
|  | H | 11 | 45.8% | 2 | 8.3% | 5 | **20.8%** | 6 | 25.0% | 24 |
|  | DH | 16 | 23.2% | 11 | 15.9% | 7 | 10.1% | 35 | **50.7%** | 69 |
|  | Total | 236 |  | 81 |  | 19 |  | 69 |  | 405 |
| M12 to M18 | NS | 170 | **78.3%** | 23 | 10.6% | 6 | 2.8% | 18 | 8.3% | 217 |
|  | D | 30 | 38.0% | 36 | **45.6%** | 3 | 3.8% | 10 | 12.7% | 79 |
|  | H | 6 | 26.1% | 3 | 13.0% | 8 | **34.8%** | 6 | 26.1% | 23 |
|  | DH | 14 | 22.2% | 10 | 15.9% | 6 | 9.5% | 33 | **52.4%** | 63 |
|  | Total | 220 |  | 72 |  | 23 |  | 67 |  | 382 |
| M18 to M24 | NS | 166 | **78.3%** | 30 | 14.2% | 3 | 1.4% | 13 | 6.1% | 212 |
|  | D | 22 | 34.4% | 32 | **50.0%** | 0 | 0.0% | 10 | 15.6% | 64 |
|  | H | 6 | 26.1% | 3 | 13.0% | 6 | **26.1%** | 8 | 34.8% | 23 |
|  | DH | 19 | 28.8% | 16 | 24.2% | 6 | 9.1% | 25 | **37.9%** | 66 |
|  | Total | 213 |  | 81 |  | 15 |  | 56 |  | 365 |

Note. Symt = Symptom; FPr = First prodromal symptom; FPs = First Psychotic Symptom; FPs1 = First Psychotic Symptom lasting at least one week; NS/O = No symptoms/Other prior to baseline; D = Only Delusions; H = Only Hallucinations; DH = Delusions and Hallucinations. Proportions were computed based on the number of participants with the symptoms from the previous visits as conditional probabilities.

**Table 21. Comparisons of transition probabilities**

| Comparisons | NAPLS 2 | | NAPLS 3 | | PEPP | |
|---|---|---|---|---|---|---|
| No symptoms (NS) | Z-Value | P-Value | Z-Value | P-Value | Z-Value | P-Value |
| NS to NS vs NS to D | 14.04 | < 0.001 | 19.05 | < 0.001 | 37.91 | < 0.001 |
| NS to NS vs NS to H | 15.49 | < 0.001 | 20.81 | < 0.001 | 43.5 | < 0.001 |
| NS to NS vs NS to HD | 16.35 | < 0.001 | 19.84 | < 0.001 | 41.39 | < 0.001 |
| NS to D vs NS to H | 2.269 | 0.023 | 3.008 | 0.0027 | 9.92 | < 0.001 |
| NS to D vs NS to HD | 3.92 | < 0.001 | 1.25 | 0.21 | 5.69 | < 0.001 |
| NS to H vs NS to HD | 1.83 | 0.067 | -1.79 | 0.074 | -4.69 | < 0.001 |
| Delusions (D) | | | | | | |
| D to NS vs D to D | -9.67 | < 0.001 | -17.33 | < 0.001 | -6.16 | < 0.001 |
| D to NS vs D to H | 7.49 | < 0.001 | 10.40 | < 0.001 | 17.25 | < 0.001 |
| D to NS vs D to HD | 1.92 | 0.055 | 3.62 | < 0.001 | 12.74 | < 0.001 |
| D to D vs D to H | 15.57 | < 0.001 | 24.50 | < 0.001 | 22.18 | < 0.001 |
| D to D vs D to HD | 11.35 | < 0.001 | 20.18 | < 0.001 | 18.23 | < 0.001 |
| D to H vs D to HD | -5.88 | < 0.001 | -7.57 | < 0.001 | -6.36 | < 0.001 |
| Hallucinations (H) | | | | | | |
| H to NS vs H to D | 5.52 | < 0.001 | 4.11 | < 0.001 | 7.26 | < 0.001 |
| H to NS vs H to H | 0.43 | 0.67 | -5.67 | < 0.001 | 3.18 | 0.0015 |
| H to NS vs H to HD | 4.2 | < 0.001 | -18.22 | < 0.001 | 2.32 | 0.021 |
| H to D vs H to H | -5.14 | < 0.001 | -9.01 | < 0.001 | -4.42 | < 0.001 |
| H to D vs H to HD | -1.55 | 0.122 | -20.4 | < 0.001 | -5.21 | < 0.001 |
| H to H vs H to HD | 3.8 | < 0.001 | -13.74 | < 0.001 | -0.87 | 0.382 |
| Hallucinations and Delusions (HD) | | | | | | |
| HD to NS vs HD to D | -3.48 | < 0.001 | -7.65 | < 0.001 | 2.03 | 0.042 |
| HD to NS vs HD to H | 3.35 | < 0.001 | -0.63 | 0.529 | 7.01 | < 0.001 |
| HD to NS vs HD to HD | -16 | < 0.001 | -32.85 | < 0.001 | -10.27 | < 0.001 |
| HD to D vs HD to H | 6.702 | < 0.001 | 7.084 | < 0.001 | 5.029 | < 0.001 |
| HD to D vs HD to HD | -12.94 | < 0.001 | -27.6 | < 0.001 | -12.18 | < 0.001 |
| HD to H vs HD to HD | -18.55 | < 0.001 | -32.5 | < 0.001 | -16.59 | < 0.001 |

Note. Comparisons between proportions from initial state to following state were performed using proportion differences. P values are not adjusted for multiple comparisons. PEPP comparisons were performed from baseline to 24 months.

**Table S22. Time in months lapsed between the onset of the first and second symptoms in CHR-P and FEP samples**

| | NAPLS 2 | | | | NAPLS 3 | | | | PEPP |
|---|---|---|---|---|---|---|---|---|---|
| | From Baseline | | Retrospective | | From Baseline | | Retrospective | | Prodrome Length |
| | Hall-Del | Del-Hall | Hall-Del | Del-Hall | Hall-Del | Del-Hall | Hall-Del | Del-Hall | First symptom to baseline |
| N | 12 | 28 | 80 | 159 | 7 | 47 | 104 | 150 | 591 |
| Mean | 10.33 | 11.35 | 22.22 | 22.05 | 8.71 | 6.02 | 23.50 | 28.34 | 22.92 |
| SD | 8.2 | 5.66 | 25.31 | 27.95 | 6.10 | 5.61 | 27.82 | 35.48 | 38.09 |
| Median | 6 | 11 | 11 | 12 | 8 | 4 | 12 | 12 | 6.97 |
| Min | 5 | 5 | 1 | 1 | 2 | 1 | 1 | 1 | 0 |
| Max | 28 | 25 | 104 | 172 | 18 | 24 | 128 | 208 | 231.2 |
| IQR | 7.5 | 8 | 26 | 23 | 6 | 6 | 18.5 | 37 | 27.46 |

Note. Hall-Dell = participants who reported retrospectively having hallucinations as first symptom and then developed delusions; Dell-Hall = participants who reported retrospectively having delusions as first symptom and then developed hallucinations; IQR = Interquartile range. For participants who reported delusions or hallucinations at baseline and later experienced the other symptom, the time of onset for the second symptom was recorded from the baseline until the first occurrence of the second symptom during the study (From Baseline). For those who reported both delusions and hallucinations at baseline, we tracked the onset of symptoms retrospectively to report the time lapse between the symptoms (Retrospective). The prodrome length was measured in PEPP sample, considering the onset of the first prodrome symptom (hallucination or delusion) to the first psychotic symptom at baseline. No significant differences in time lapsed between hallucinations and delusions were found.

**Table S23. Lempel Ziv Complexity index for hallucination and delusions**

|  | NAPLS 2 | | NAPLS 3 | | PEPP | |
|---|---|---|---|---|---|---|
|  | Del | Hall | Del | Hall | Del | Hall |
| N | 383 | 337 | 301 | 260 | 598 | 473 |
| Mean | 1.22 | 1.29 | 1.06 | 1.11 | 1.24 | 1.25 |
| SEM | 0.01 | 0.01 | 0.02 | 0.02 | 0.01 | 0.01 |
| Median | 1.05 | 1.39 | 1.13 | 1.13 | 1.20 | 1.20 |
| Mode | 1.39 | 1.39 | 0.75 | 1.20 | 1.05 | 1.05 |
| SD | 0.27 | 0.56 | 0.27 | 0.29 | 0.27 | 0.26 |
| Minimum | 0.92 | 0.92 | 0.75 | 0.70 | 0.70 | 0.70 |
| Maximum | 1.85 | 1.87 | 2.00 | 1.85 | 2.00 | 2.01 |

Note. LZC values are standardized values.

For the purposes of this study, the presence or absence of hallucinations and delusions in the prospective sample was established using the SAPS questions 7 (global assessment of hallucinations) and 20 (global assessment of delusions); scores of 0 (none) and 1 (questionable) were coded as 'absent' and scores of 2 (mild) and above were coded as present. While symptomatic remission of symptoms on the SAPS is often considered to be present at mild symptoms and below,[15] our choice for binarization was made as we were interested in following subtle changes in symptoms over time that might reflect underlying computational processes. In addition, as this was a treated sample, examining mild symptoms allowed for the assessment of symptoms that were present though symptomatically controlled by medication, which is likely relevant to understanding symptom transitions over time. In Table S24, we provide the data on symptom prevalence at each timepoint using the clinical cutoff of 3 points. As can be seen, the patterns observed- with delusions, and delusions and hallucinations being more common than hallucinations- are maintained from the analysis with the alternative cutoff.

**Table S24. PEPP. Frequency and percentage of symptoms from baseline PEPP admission to 24 months using the 3 or greater as a cutoff on the SAPS items**

| Time | | NS/O | Del | Hall | Del&Hall | Total |
|---|---|---|---|---|---|---|
| BL | Fr | 29 | 242 | 21 | 385 | 677 |
|    | %  | 4.3 | 35.7 | 3.1 | 56.9 | 100 |
| M1 | Fr | 184 | 97 | 32 | 65 | 378 |
|    | %  | 48.7 | 25.7 | 8.5 | 17.2 | 100 |
| M2 | Fr | 318 | 77 | 28 | 55 | 478 |
|    | %  | 66.5 | 16.1 | 5.9 | 11.5 | 100 |
| M3 | Fr | 410 | 76 | 37 | 56 | 563 |
|    | %  | 70.8 | 13.1 | 6.4 | 9.7 | 100 |
| M6 | Fr | 386 | 70 | 32 | 48 | 536 |
|    | %  | 72 | 13.1 | 6 | 9 | 100 |
| M9 | Fr | 339 | 50 | 32 | 37 | 458 |
|    | %  | 74 | 10.9 | 7 | 8.1 | 100 |
| M12 | Fr | 360 | 42 | 24 | 48 | 474 |
|     | %  | 75.9 | 8.9 | 5.1 | 10.1 | 100 |
| M18 | Fr | 319 | 39 | 33 | 38 | 429 |
|     | %  | 74.4 | 9.1 | 7.7 | 8.9 | 100 |
| M24 | Fr | 314 | 45 | 13 | 39 | 411 |
|     | %  | 76.4 | 10.9 | 3.2 | 9.5 | 100 |

Note. NS= No symptoms (score of absent, questionable, or mild on SAPS item); Del = Delusions only; Hall = Hallucinations only; Del&Hall = Delusions and hallucinations

**Table S25. NAPLS 2, Total symptom severity, Perceptual Abnormalities (P4), Unusual Thought Content (P1), and Suspiciousness or Persecutory Ideas (P2)**

|  |  | Total Scale of Psychosis-risk Symptoms (SOPS) | | | | | Perceptual Abnormalities and Hallucinations (P4) | | | | | Unusual Thought Content (P1) & Suspiciousness or Persecutory Ideas (P2) | | | | |
|---|---|---|---|---|---|---|---|---|---|---|---|---|---|---|---|---|
|  |  | BL | 6M | 12M | 18M | 24M | BL | 6M | 12M | 18M | 24M | BL | 6M | 12M | 18M | 24M |
| Hall-Del | N | 76 | 58 | 40 | 32 | 32 | 80 | 58 | 40 | 32 | 32 | 80 | 58 | 40 | 32 | 32 |
|  | M | 39.5 | 30.1 | 27 | 24.7 | 22 | 3.9 | 2.8 | 2.1 | 1.6 | 1.8 | 3.7 | 3.1 | 2.8 | 2.3 | 2.3 |
|  | SD | 11.3 | 13.2 | 11.4 | 11.5 | 12.2 | 0.7 | 1.5 | 1.4 | 1.4 | 1.4 | 0.7 | 1.4 | 1.5 | 1.5 | 1.5 |
| Del-Hall | N | 156 | 107 | 90 | 59 | 59 | 158 | 108 | 90 | 59 | 59 | 158 | 108 | 90 | 59 | 59 |
|  | M | 42.2 | 32.5 | 28.6 | 28.1 | 26.1 | 3.6 | 2.6 | 2.2 | 2 | 1.8 | 4.1 | 3.4 | 3 | 2.8 | 2.8 |
|  | SD | 11.4 | 13.7 | 13.5 | 14.1 | 12.2 | 0.7 | 1.6 | 1.6 | 1.6 | 1.5 | 0.7 | 1.1 | 1.3 | 1.4 | 1.3 |
| Del&Hall | N | 262 | 177 | 153 | 103 | 102 | 275 | 178 | 153 | 104 | 102 | 275 | 178 | 153 | 104 | 102 |
|  | M | 38.2 | 28.8 | 25.8 | 23.5 | 25.4 | 3.9 | 2.6 | 2.4 | 2 | 1.9 | 3.9 | 3 | 2.8 | 2.5 | 2.3 |
|  | SD | 11 | 13.3 | 12.3 | 13.9 | 15.4 | 0.7 | 1.5 | 1.6 | 1.6 | 1.7 | 0.8 | 1.4 | 1.4 | 1.6 | 1.6 |
| Del | N | 160 | 110 | 85 | 49 | 53 | 165 | 110 | 85 | 49 | 53 | 165 | 110 | 85 | 49 | 53 |
|  | M | 37.1 | 28 | 25 | 24 | 23.2 | 1 | 0.9 | 1 | 1.1 | 0.8 | 3.9 | 2.9 | 2.5 | 2.5 | 2.5 |
|  | SD | 11.7 | 12.9 | 13.4 | 13.8 | 13.4 | 0.9 | 1.3 | 1.4 | 1.3 | 1.2 | 0.8 | 1.3 | 1.3 | 1.2 | 1.3 |
| Hall | N | 39 | 33 | 25 | 18 | 17 | 40 | 33 | 25 | 18 | 17 | 40 | 33 | 25 | 18 | 17 |

|   | M | 26.1 | 19.5 | 19 | 16.9 | 16 | 3.8 | 2 | 1.6 | 1.6 | 1.4 | 1.2 | 1.7 | 1.3 | 1.1 | 1.1 |
|   | SD | 12.4 | 11.7 | 11.4 | 11.7 | 9.6 | 0.7 | 1.6 | 1.5 | 1.8 | 1.4 | 0.8 | 1.4 | 1.2 | 1.4 | 1.2 |

Note. Hall-Dell = participants who reported retrospectively having hallucinations as first symptom and then developed delusions; Dell-Hall = participants who reported retrospectively having delusions as first symptom and then developed hallucinations; Del&Hall = participants who reported delusions and hallucinations at the same onset; Del = participants that reported delusions as a first symptom; Hall = participants who reported hallucinations as a first symptom.

**Table S26. NAPLS 2, Disorganized, Negative, General, and Positive Symptoms.**

|  | SOPS Disorganized Symptoms | | | SOPS Negative Symptoms | | | SOPS Positive Symptoms | | | SOPS General Symptoms | | |
|---|---|---|---|---|---|---|---|---|---|---|---|---|
|  | N | M | SD | N | M | SD | N | M | SD | N | M | SD |
| Hall-Del | 76 | 5.00 | 2.766 | 76 | 12.20 | 5.683 | 80 | 12.35 | 2.605 | 76 | 10.04 | 4.165 |
| Del-Hall | 156 | 5.53 | 3.076 | 156 | 13.06 | 5.964 | 159 | 13.69 | 2.955 | 156 | 9.96 | 4.258 |
| Del&Hall | 263 | 5.10 | 3.158 | 262 | 11.14 | 5.825 | 275 | 12.86 | 3.197 | 262 | 9.13 | 4.010 |
| Hall | 39 | 4.38 | 3.361 | 39 | 8.64 | 6.372 | 40 | 7.18 | 2.678 | 39 | 6.00 | 4.668 |
| Del | 161 | 5.17 | 3.067 | 161 | 12.66 | 6.101 | 165 | 10.46 | 3.247 | 160 | 8.84 | 4.262 |
| Total | 695 | 5.16 | 3.092 | 694 | 11.90 | 6.026 | 719 | 12.12 | 3.482 | 693 | 9.17 | 4.265 |

Note. Hal-Dell = participants who reported retrospectively having hallucinations as first symptom and then developed delusions; Dell-Hall = participants who reported retrospectively having delusions as first symptom and then developed hallucinations; Del&Hall = participants who reported delusions and hallucinations at the same onset; Del = participants that reported delusions as a first symptom; Hall = participants who reported hallucinations as first symptoms. No significant differences were found for Disorganized symptoms. Negative symptoms ($F_{4,689} = 6.232$, $P < 0.001$, $\eta^2 = .035$, Hall = Del&Hall, Hall < Del, Del-Hall and Hall-Del $Ps < 0.05$) ; Positive symptoms ($F_{4,714} = 16.440$, $P < 0.001$, $\eta^2 = .035$, Hall < all groups, Dell < Del-Hall, Del&Hall, Hall-Del); General symptoms ($F_{4,688} = 8.085$, $P < 0.001$, $\eta^2 = .045$, Hall < all groups).

**Table S27. NAPLS 3, Total symptom severity, Perceptual Abnormalities (P4), Unusual Thought Content (P1), and Suspiciousness or Persecutory Ideas (P2)**

| | | Total scores for Scale of Psychosis-risk Symptoms (SOPS) | | | | | | | | Perceptual Abnormalities and Hallucinations (P4) | | | | | | | | | Unusual Thought Content (P1) & Suspiciousness or Persecutory Ideas (P2) | | | | | | | |
|---|---|---|---|---|---|---|---|---|---|---|---|---|---|---|---|---|---|---|---|---|---|---|---|---|---|---|
| | | BL | 2M | 4M | 6M | 8M | 12M | 18M | 24M | BL | 2M | 4M | 6M | 8M | 12M | 18M | 24M | BL | 2M | 4M | 6M | 8M | 12M | 18M | 24M |
| Hall-Del | N | 105 | 56 | 52 | 51 | 46 | 49 | 38 | 32 | 105 | 75 | 70 | 62 | 57 | 62 | 46 | 41 | 105 | 75 | 70 | 62 | 57 | 62 | 46 | 41 |
| | M | 38.7 | 34.3 | 29.6 | 28.8 | 26.1 | 25.8 | 23.8 | 21.6 | 3.9 | 3.3 | 3 | 2.8 | 2.5 | 2.4 | 2 | 1.8 | 4 | 3.4 | 3.2 | 3 | 2.8 | 2.8 | 2.4 | 2.2 |
| | SD | 12.1 | 12.3 | 13.9 | 14.3 | 12.6 | 13.1 | 13.3 | 11.7 | 0.7 | 1.2 | 1.5 | 1.5 | 1.6 | 1.5 | 1.6 | 1.5 | 0.6 | 1.2 | 1.5 | 1.2 | 1.4 | 1.4 | 1.6 | 1.5 |
| Del - Hall | N | 153 | 84 | 71 | 72 | 73 | 65 | 55 | 48 | 154 | 103 | 85 | 86 | 86 | 77 | 65 | 58 | 154 | 103 | 85 | 86 | 86 | 77 | 65 | 58 |
| | M | 41.5 | 37.9 | 36.4 | 32 | 31.1 | 29.5 | 28.7 | 28.5 | 3.7 | 3.1 | 2.8 | 2.7 | 2.3 | 1.8 | 1.7 | 1.6 | 4.3 | 3.8 | 3.6 | 3.2 | 3.1 | 2.9 | 2.7 | 2.7 |
| | SD | 12.1 | 12.3 | 13.7 | 14.1 | 14.5 | 14.2 | 13.6 | 14.5 | 0.7 | 1.3 | 1.5 | 1.4 | 1.4 | 1.4 | 1.4 | 1.3 | 0.6 | 1 | 1.2 | 1.3 | 1.3 | 1.2 | 1.3 | 1.2 |
| Del & Hall | N | 236 | 129 | 128 | 111 | 111 | 96 | 85 | 75 | 240 | 163 | 151 | 129 | 135 | 115 | 97 | 85 | 240 | 163 | 151 | 129 | 135 | 115 | 97 | 85 |
| | M | 40.4 | 35.5 | 33.1 | 32.9 | 29.4 | 27.6 | 24.8 | 26.3 | 3.8 | 3.3 | 2.9 | 2.6 | 2.5 | 2.2 | 2.1 | 2 | 4.2 | 3.8 | 3.6 | 3.2 | 3 | 2.8 | 2.5 | 2.5 |
| | SD | 12.7 | 13.7 | 14.3 | 14.9 | 16.6 | 15 | 15 | 15.3 | 0.7 | 1.2 | 1.5 | 1.6 | 1.7 | 1.6 | 1.5 | 1.6 | 0.7 | 1 | 1.2 | 1.4 | 1.5 | 1.5 | 1.5 | 1.5 |
| Del | N | 177 | 95 | 82 | 81 | 79 | 75 | 60 | 42 | 179 | 123 | 112 | 102 | 103 | 97 | 71 | 51 | 179 | 123 | 112 | 102 | 103 | 97 | 71 | 51 |
| | M | 38.8 | 34.4 | 33.3 | 29.2 | 29.7 | 28 | 26.9 | 23.6 | 1.2 | 1.2 | 1.1 | 1.1 | 1.2 | 1.1 | 1.1 | 0.8 | 4.1 | 3.4 | 3.1 | 2.9 | 2.7 | 2.6 | 2.5 | 2.3 |
| | SD | 13 | 14.9 | 14.9 | 14.1 | 13.7 | 16.7 | 15.3 | 14 | 0.9 | 1.2 | 1.2 | 1.3 | 1.3 | 1.3 | 1.2 | 1 | 0.7 | 1.2 | 1.3 | 1.4 | 1.4 | 1.5 | 1.5 | 1.5 |
| Hall | N | 16 | 1 | 2 | 2 | 1 | 2 | 0 | 0 | 21 | 19 | 17 | 17 | 16 | 16 | 9 | 9 | 21 | 19 | 17 | 17 | 16 | 16 | 9 | 9 |
| | M | 24.4 | 12 | 15 | 10 | 9 | 15.5 | 0 | 0 | 4 | 3.5 | 2.8 | 2.9 | 2.6 | 2.4 | 2.7 | 1.8 | 1.6 | 1.6 | 1.4 | 1.2 | 1.4 | 1.7 | 1.7 | 1.1 |
| | SD | 11.5 | 0 | 8.5 | 0 | 0 | 2.1 | 0 | 0 | 0.6 | 1 | 1.5 | 1.5 | 1.2 | 1.6 | 1.6 | 1.6 | 0.5 | 0.8 | 0.9 | 1 | 1.2 | 1.6 | 1.7 | 1.3 |

Note. Hall-Dell = participants who reported retrospectively having hallucinations as first symptom and then developed delusions; Dell-Hall = participants who reported retrospectively having delusions as first symptom and then developed hallucinations; Del&Hall = participants who reported delusions and hallucinations at the same onset; Del = participants that reported delusions as a first symptom; Hall = participants who reported hallucinations as a first symptom.

**Table S28. NAPLS 3, Disorganized, Negative, and Positive Symptoms.**

|          | SOPS Disorganized Symptoms | | | SOPS Negative Symptoms | | | SOPS Positive Symptoms | | | SOPS General Symptoms | | |
|----------|-----|------|-------|-----|-------|-------|-----|-------|-------|-----|------|-------|
|          | N   | M    | SD    | N   | M     | SD    | N   | M     | SD    | N   | M    | SD    |
| Hall-Del | 105 | 4.94 | 2.905 | 105 | 10.91 | 5.908 | 105 | 13.13 | 3.01  | 105 | 9.71 | 4.12  |
| Del-Hall | 153 | 4.91 | 3.042 | 154 | 12.94 | 6.179 | 154 | 14.05 | 2.846 | 153 | 9.65 | 4.156 |
| Del&Hall | 237 | 5.26 | 3.084 | 236 | 11.61 | 6.08  | 240 | 13.74 | 3.241 | 236 | 9.84 | 4.298 |
| Hall     | 16  | 2.44 | 1.75  | 16  | 7.81  | 6.102 | 21  | 7.86  | 2.535 | 16  | 6.06 | 4.234 |
| Del      | 177 | 5.59 | 3.367 | 177 | 12.94 | 6.562 | 179 | 11.44 | 3.248 | 177 | 8.8  | 4.161 |
| Total    | 688 | 5.15 | 3.132 | 688 | 12.05 | 6.272 | 699 | 12.95 | 3.384 | 687 | 9.42 | 4.244 |

Note. Hall-Del = participants who reported retrospectively having hallucinations as first symptom and then developed delusions; Del-Hall = participants who reported retrospectively having delusions as first symptom and then developed hallucinations; Del&Hall = participants who reported delusions and hallucinations at the same onset; Del = participants that reported delusions as a first symptom; Hall = participants who reported hallucinations as first symptoms. Disorganized symptoms ($F_{4,689} = 4.360$, $P = 0.002$, $\eta^2 = .025$, Hall < all groups, $Ps < 0.05$); Negative symptoms ($F_{4,683} = 4.748$, $P < 0.001$, $\eta^2 = .027$, Hall < Del-Hall and Del, $Ps < 0.05$); Positive symptoms ($F_{4,694} = 33.528$, $P < 0.001$, $\eta^2 = .162$, Hall < all groups, Del < Del-Hall, Del&Hall, Hall-Del, $Ps < 0.05$); General symptoms ($F_{4,714} = 4.365$, $P = 0.002$, $\eta^2 = .025$, Hall < all groups, Del < Del-Hall).

**Table S29. Presence prior to prodrome and course of Delusion and Hallucination-Spectrum symptoms based on the TOPE**

|  | Odd/Bizarre Ideas (n = 431) | Suspiciousness/ ideas of reference (n = 431) | Delusions (n = 343) | Unusual Perceptual Experiences (n = 434) | Hallucination (n = 344) |
|---|---|---|---|---|---|
| Ever experienced prior to psychosis onset? | 131 (30.4%) | 168 (39%) | 16 (4.7%) | 78 (18%) | 18 (5.2%) |
|  | Course (n= 389) | Course (n = 386) | Course (n = 11) | Course (n = 406) | Course (n = 14) |
| Not present during prodrome | 176 (45.2%) | 97 (25.1%) | Missing | 218 (53.7%) | Missing |
| Present during one phase | 20 (5.1%) | 28 (7.3%) | 8 (72.7%) | 14 (3.4%) | 7 (50%) |
| Present during several phases | 14 (3.6%) | 28 (7.3%) | 1 (9.1%) | 18 (4.4%) | 6 (42.9%) |
| Persistent | 54 (13.9%) | 81 (21%) | None | 30 (7.4%) | None |
| Lifelong pattern of behavior | 11 (2.8%) | 7 (1.8%) | None | 3 (0.7%) | None |
| Uncertain | 18 (4.6%) | 13 (3.4%) | 2 (18.2%) | 23 (5.7%) | 1 (7.1%) |
| Only after psychosis onset | 96 (24.7%) | 132 (34.2%) | Missing | 100 (24.6%) | Missing |

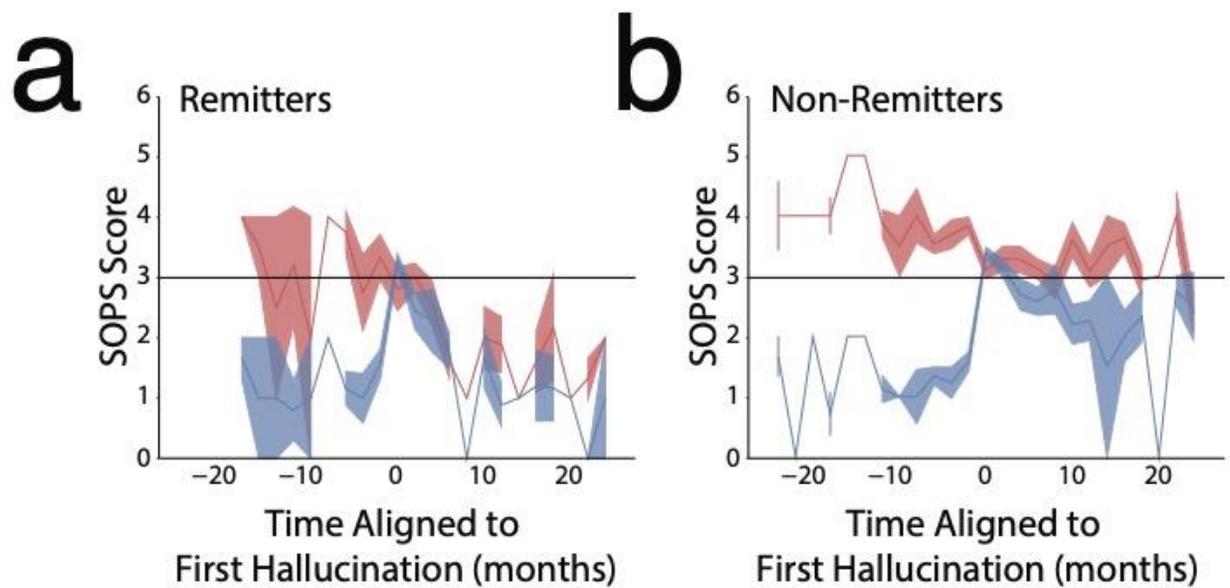

**Supplemental Figure S1. Combined NAPLS 2 and 3 symptom onset trajectories, split by remission status.** In both participants whose symptoms remitted (left) and those whose symptoms persisted or got worse over time (right), hallucination onset was preceded by a period of elevated delusional intensity. After hallucination onset, both remitters and non-remitters demonstrated a decrease in hallucination and delusion intensity. Red = hallucinations; blue = delusions.

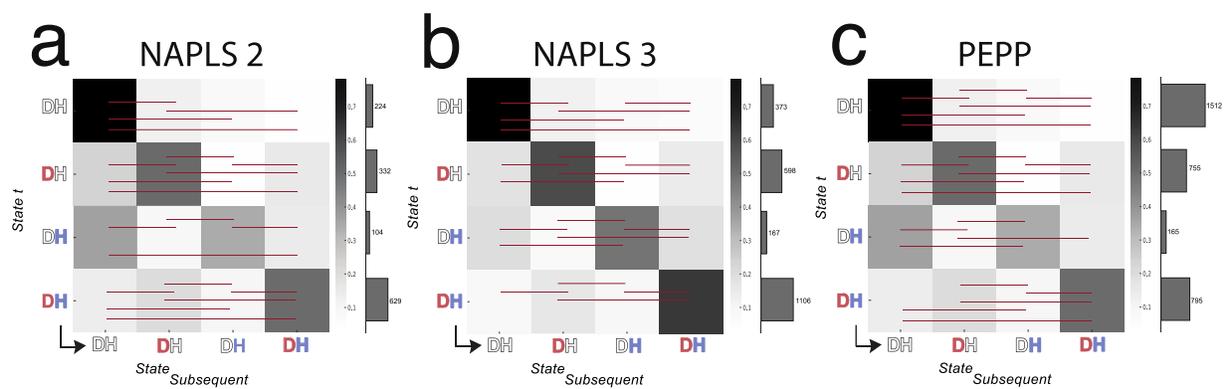

**Supplemental Figure S2. Full transition matrices** from symptom states (y) to immediately subsequent symptom states across all three samples. Comparisons reported in **Table S21.**

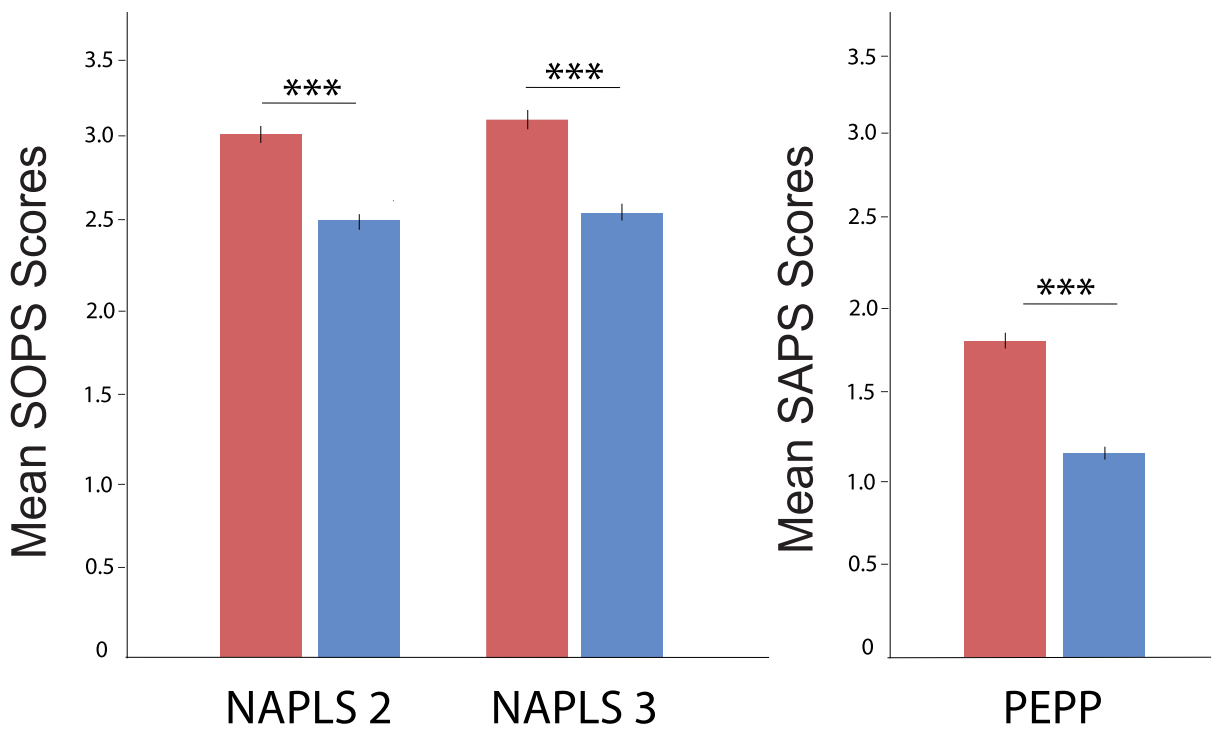

**Supplemental Figure S3. Symptom severity in NAPLS 2, NAPLS 3, and PEPP samples.**
Mean scores for P4, P1, and P2 over time are presented, with sample selection criteria identical to the LZC analysis. Error bars = 1 SEM.